# The ZZ feature map induces a signless Laplacian metric: a closed-form classical surrogate for quantum kernel regression

Erkut Tekeli[1,*]
[1] Department of Software Engineering, Muğla Sıtkı Koçman University, Muğla, Türkiye

Contact author: erkuttekeli@mu.edu.tr

**ABSTRACT.** Quantum kernel methods have been reported to lose their advantage over classical kernels once the encoding bandwidth is tuned, and bandwidth-tuned quantum kernels have been shown to resemble radial basis function kernels closely. The analytical support for that observation, however, rests on separable encoding circuits, and its authors note that it captures entangling circuits only qualitatively. We close this gap for the ZZ feature map. We prove that in the small-bandwidth regime the induced kernel has, to leading order, the anisotropic Mahalanobis geometry defined by $M = I + \pi^2 Q$, where $Q$ is the signless Laplacian of the entanglement graph, and that it admits a parameter-free Gaussian surrogate matching it through second order. The quadratic structure persists at any fixed depth as a pullback of the Fubini–Study metric. We also show that the anisotropy depends entirely on a phase convention: under the unshifted convention, the metric is the identity irrespective of entanglement, which supplies an analytical account of the isotropic resemblance previously reported. The derivation is verified against direct simulation for path, cycle, and complete entanglement graphs, with relative error below $10^{-3}$. Building the corresponding classical kernel (which requires no fitted parameters and no quantum simulation, since the metric is read off the entanglement graph), we compare it with the quantum kernel on two near-infrared spectroscopic regression benchmarks, across four targets, five preprocessing pipelines, and 100 resampled splits per cell. The paired 95% bootstrap interval contains zero in 18 of 20 cells, the typical relative difference in test error is 2.7%, and the two families select the same preprocessing pipeline in 96% of splits. The regime in which the reduction fails begins at the same bandwidth on both datasets and adds no robust predictive value: restricting the grid to the classical regime improves mean test error by 3.8%, a gain that an equally large random restriction does not reproduce. In this configuration, the quantum circuit can be removed from the pipeline without detectable predictive loss, and we can identify the metric it induces and a parameter-free classical kernel that reproduces it to leading order.

## I. INTRODUCTION

Quantum kernel methods occupy an unusual position in quantum machine learning. Unlike variational models, whose training dynamics are difficult to characterise, a quantum kernel method is a classical convex learning algorithm that happens to evaluate its Gram matrix on quantum hardware (Havlíček et al., 2019; Schuld & Killoran, 2019). The quantum device contributes exactly one thing: a similarity measure induced by a data-encoding circuit. This modularity is attractive both practically, because the surrounding statistics are standard and well understood, and theoretically, because any advantage must be attributable to the encoding map alone. If a quantum feature map yields better generalisation than every classical kernel on a given task, the reason must lie in the geometry it induces on the input space.

The observation that many supervised quantum models admit a kernel-method formulation (Schuld, 2021; Schuld & Killoran, 2019) places the data-encoding circuit at the centre of the analysis. For the corresponding kernel model, optimization of the classifier is classical and convex; the encoding determines the quantum-induced similarity measure and hence the model's inductive bias. Related constructions differ in the kernel estimator or classifier layered on top of it (Blank et al., 2020; Park et al., 2020), while work on data encoding approaches the same issue from the variational side: data encodings determine the Fourier structure of represented functions (Schuld et al., 2021), and embeddings themselves can be optimized as trainable objects (Lloyd et al., 2020).

A series of results has clarified the restrictive conditions under which a quantum kernel can outperform classical alternatives. Huang et al. (2021) showed that access to training data can substantially reduce an apparent quantum advantage and introduced the geometric difference as a function-independent diagnostic: a small geometric difference with respect to an efficient classical model implies that the classical model can match or improve the quantum model's predictive performance. Huang, Kueng, and Preskill (2021) derived complementary information-theoretic bounds on quantum advantages in learning. On the positive side, Liu, Arunachalam, and Temme (2021) constructed a supervised classification problem based on the assumed classical hardness of discrete logarithms. They proved an end-to-end, noise-robust quantum speed-up for this constructed problem with classical access to the data; it should therefore not be read as evidence for generic classical-data advantages. Jerbi et al. (2023) identify learning tasks for which data re-uploading models can be more resource-efficient than linear quantum models and kernel methods, while Cerezo et al. (2022) survey the broader opportunities and limitations of quantum machine learning.

A parallel line of work asks not whether a formal separation is possible, but whether highly expressive encodings provide a useful inductive bias. Kübler et al. (2021) characterize the inductive bias of quantum kernels and relate poor generalization to spectral properties of the induced kernel. Gil-Fuster, Eisert, and Dunjko (2024) study the expressivity of embedding quantum kernels directly. Thanasilp et al. (2024) show exponential concentration of fidelity-kernel values for broad classes of expressive encodings: as the system size or circuit expressivity grows, kernel values can concentrate around a constant, impairing discrimination and trainability. Alternative kernel constructions, including quantum Fisher kernels, have been proposed to mitigate vanishing-similarity behavior under specified ansätze (Suzuki, Kawaguchi, & Yamamoto, 2024); such proposals do not remove the need to analyze the particular encoding and measurement scheme used.

A direct response to concentration is to rescale the classical data before encoding. Shaydulin and Wild (2022) establish quantum-kernel bandwidth as a central hyperparameter, showing numerically that large bandwidths can underfit, small bandwidths can overfit, and intermediate values can optimize generalization; bandwidth selection can also mitigate exponential decay of kernel values with the number of qubits. Canatar et al. (2023) provide an analytical generalization theory in which bandwidth controls the kernel spectrum and inductive bias, and can change a model from one that provably cannot generalize to one that generalizes for suitably aligned targets. Haug, Self, and Kim (2023) address measurement-efficient estimation of quantum kernels for large datasets.

Under carefully controlled comparisons, Slattery et al. (2023) found no evidence of an advantage for the fidelity kernels they studied on classical datasets. Peters et al. (2021) reported the practical difficulties of learning high-dimensional data on noisy quantum hardware. Bowles, Ahmed, and Schuld (2024) emphasize that conclusions depend strongly on benchmark design, including classical baselines, preprocessing, hyperparameter optimization, and ablation choices. Their large-scale study found that their out-of-the-box classical baselines systematically outperformed the examined quantum models on the small-scale datasets considered; this is a methodological result for the tested setting, not a universal impossibility statement.

A smaller body of work designs the encoding to reflect known data structure rather than maximizing generic circuit expressivity. Glick et al. (2024) introduce covariant quantum kernels for data carrying group structure, and Henry et al. (2021) introduce a quantum evolution kernel for graph learning using programmable qubit arrays. These works are conceptually close to approaches that treat the geometry induced by the circuit as an object for analysis rather than as a black box.

Flórez-Ablan, Roth, and Schnabel (2025) study similarities between bandwidth-tuned quantum kernels and classical kernels across several encoding circuits and datasets. They report that the optimized kernels can be well approximated by low-order Taylor expansions of an RBF kernel in the small-bandwidth regime observed in their experiments. Their analytical model is derived for a separable single-qubit-rotation encoding whose kernel factorizes across coordinates. The authors describe the extension to entangling circuits as qualitative rather than exact. Thus, their analysis does not settle the behavior of entangling feature maps such as ZZ/IQP-style maps, for which pairwise phases change the geometry of the induced kernel.

The signless Laplacian $Q = D + A$ is a standard object in spectral graph theory (Cvetković, Rowlinson, & Simić, 2007; Cvetković & Simić, 2009; Cvetković, Rowlinson, & Simić, 2010). For a vector $u$, its quadratic form is

$$\boldsymbol{u}^{\top} Q \boldsymbol{u} = \sum_{\{i,j\}\in E} (u_i + u_j)^2.$$

We also use the classical characterization that the least signless-Laplacian eigenvalue is zero precisely when the graph has a bipartite connected component (Desai & Rao, 1994).

This paper closes that gap for the ZZ feature map. We show that in the small-bandwidth regime (which is the regime selected by cross-validation in our experiments) the induced kernel is not merely *similar to* a classical kernel but has, to leading order, exactly the anisotropic Mahalanobis geometry of a Gaussian kernel whose metric we give in closed form. The metric is determined entirely by the entanglement graph, and it is a Mahalanobis form built from that graph's signless Laplacian. We then construct the corresponding classical kernel explicitly, without fitting the graph metric and without quantum simulation, and show that it matches the quantum kernel's predictive performance on spectroscopic regression benchmarks.

Our contributions are as follows.

1. A closed-form induced metric. For the ZZ feature map with one layer and an arbitrary entanglement graph $G$, we prove that $1 - K(\mathbf{x}, \mathbf{x}') = s^2\, \mathbf{u}^{\top} M \mathbf{u} + \mathcal{O}(s^3)$ with $M = I + \pi^2 Q$, where $Q$ is the signless Laplacian of $G$ and $\mathbf{u} = \mathbf{x} - \mathbf{x}'$. The derivation follows from the orthonormality of parity functions and requires no approximation beyond the bandwidth expansion. We verify it numerically for path, cycle, and complete entanglement graphs, with relative error below $10^{-3}$.

2. Persistence across circuit depth. We show that the quadratic structure is not an artefact of the single-layer case. For any depth $L$, the encoded state is affine in the data to first order in the bandwidth, so the fidelity reduces to the pullback of the Fubini–Study metric and remains a quadratic form; the single-layer result is recovered as a special case. The metric is no longer a function of $Q$ alone for $L \geq 2$, and we determine it numerically.

3. The role of the phase convention. Two conventions for the pair phase are in circulation, differing by a shift. Writing $\mathbf{y}$ for the vector the circuit actually encodes (so that $\mathbf{y} = s\mathbf{x}$ once the bandwidth is applied), the pair phase is taken either as $\varphi_{ij}(\mathbf{y}) = (\pi - y_i)(\pi - y_j)$, as in Qiskit ZZFeatureMap, or as the plain product $\varphi_{ij}(\mathbf{y}) = y_i y_j$. We prove that under the latter, the entangling term enters only at second order in the bandwidth, so that $M = I$ and the small-bandwidth kernel is isotropic regardless of the entanglement pattern, with an error one order smaller than in the shifted case. The anisotropy of the Qiskit convention arises entirely from its $\pi$-shift. This supplies an analytical account of the isotropic-RBF resemblance reported in the literature rather than contradicting it.

4. Predictive equivalence of an explicit classical twin. We evaluate the surrogate kernel against the quantum kernel on two near-infrared spectroscopic regression benchmarks, across four targets and five preprocessing pipelines, with 100 resampled splits per cell. The paired 95% bootstrap interval for the difference in test error contains zero in 18 of 20 cells; the two exceptions lie in opposite directions. The quantum win rate is 0.46, and the typical relative difference in test error is 2.7%. The residual deviation grows with the selected bandwidth, consistent with the $\mathcal{O}(s^3)$ error term of the theory.

5. The non-classical regime adds no robust predictive value. The quadratic reduction is exact to numerical precision for $s \lesssim 7{\times}10^{-3}$ and remains above $R^2 = 0.99$ up to $s = 0.072$ and degrades above it; remarkably, this boundary is identical on both datasets, consistent with the boundary being a property of the circuit rather than of the data. We then show that the non-classical region adds no robust predictive value. Restricting the bandwidth grid to the classical regime does not degrade test error: over 2000 cells it *improves* mean test error by 3.8% (95% percentile bootstrap interval: 2.2–5.6%), the effect is monotone across six candidate thresholds, and it holds in 16 of 20 cells. Cross-validation selects the non-classical regime in 15.7% of cells, and in exactly those cells the restriction reduces test error by 17%. A control experiment rules out the obvious alternative explanation: removing an equally large random subset of bandwidths changes test error by only 0.25% with a 5–95% range covering zero, so the gain cannot be attributed to reduced selection overfitting. The one exception, discussed in Section 4, is the Tecator protein target.

Together, these results suggest that, for this family of encodings on this class of data, the quantum circuit can be removed from the computational pipeline without detectable predictive loss in our experiments and that the regime in which the kernel does compute something a classical kernel does not is, in our data, one in which it computes nothing useful. We regard this as a positive statement rather than a negative one: knowing *which* classical kernel a feature map implements is more informative than knowing only that some classical kernel performs comparably, and the graph-theoretic form of the metric makes the inductive bias of the encoding explicit and adjustable.

We also note, in passing, that fixing classical length-scale grids independently of the data placed our classical baselines at the grid boundary in every selection; Section 4.6 discusses the consequences.

The remainder of the paper is organised as follows. Section 2 introduces the feature map, derives the induced metric and its consequences, defines the classical surrogate, and describes the datasets, kernel families and evaluation protocol. Section 3 reports the numerical verification of the derivation, the boundary of the classical regime, the equivalence experiments and the benchmark comparison. Section 4 discusses the implications for reported quantum kernel advantages and for hardware studies of these circuits; Section 5 states the limitations of the analysis, and Section 6 concludes.

## II. METHODS

### A. The feature map and its induced metric

#### *1. Setup and notation*

Let $\mathbf{x} \in \mathbb{R}^k$ denote a preprocessed input vector and let $s > 0$ be a global bandwidth (scaling) parameter, so that the circuit is evaluated at $s\mathbf{x}$. We use one qubit per feature, $k$ qubits in total. The entanglement structure is described by a simple undirected graph $G = (V, E)$ with $V = \{1, \dots, k\}$; the linear, circular, and full entanglement patterns correspond respectively to the path $P_k$, the cycle $C_k$ , and the complete graph $K_k$.

A single layer of the feature map applies a Hadamard on every qubit followed by a diagonal phase unitary,

$$U(\mathbf{y}) = D(\mathbf{y})\, H^{\otimes k}, \qquad (1)$$

where, with the convention $R_Z(\theta) = \exp(-i\theta Z/2)$,

$$D(\mathbf{y}) = \exp\left(-i\left[\sum_{i\in V} y_i\, Z_i + \sum_{(i,j)\in E} \varphi_{ij}\,(\mathbf{y})\, Z_i Z_j\right]\right). \qquad (2)$$

Each two-qubit term is realised in the usual way by a CX–$R_Z$–CX block. The full map with $L$ layers is $|\phi(\mathbf{y})\rangle = \left[D(\mathbf{y})H^{\otimes k}\right]^L |0\rangle^{\otimes k}$, and the fidelity kernel is

$$K(\mathbf{x}, \mathbf{x}') = |\langle \phi(s\mathbf{x}) \mid \phi(s\mathbf{x}')\rangle|^2\,. \qquad (3)$$

Two conventions for the pair phase appear in the literature. We write

$$\varphi_{ij}^{ZZ}(\mathbf{y}) = (\pi - y_i)\big(\pi - y_j\big), \qquad \varphi_{ij}^{IQP}(\mathbf{y}) = y_i y_j, \qquad (4)$$

the first being the default of the ZZFeatureMap in Qiskit and the second the form used in much of the analytical literature. As we show below, the distinction is not cosmetic: it determines whether the induced metric is anisotropic or isotropic.

It is convenient to index the computational basis by sign vectors. For $b \in \{0,1\}^k$ put $z_i = (-1)^{b_i}$, so $z \in \{-1,1\}^k$, and let $\mathbb{E}_z[\cdot] = 2^{-k} \sum_z [\,\cdot\,]$ denote the uniform expectation. The parity functions $\chi_S(z) = \prod_{i\in S} z_i$, $S \subseteq V$, form an orthonormal system,

$$\mathbb{E}_z[\chi_S \chi_T] = \delta_{S,T}, \qquad \mathbb{E}_z[\chi_S] = \delta_{S,\varnothing}. \qquad (5)$$

Since $D(\mathbf{y})$ is diagonal, $D(\mathbf{y})|z\rangle = e^{-if_z(\mathbf{y})}|z\rangle$ with

$$f_z(\mathbf{y}) = \sum_{i\in V} y_i\, z_i + \sum_{(i,j)\in E} \varphi_{ij}(\mathbf{y})\, z_i z_j. \qquad (6)$$

***2. The small-bandwidth expansion for a single layer***

For $L = 1$ we have $|\,\phi(\mathbf{y})\rangle = 2^{-k/2}\sum_z e^{-if_z(\mathbf{y})}\,|\,z\rangle$ and therefore

$$K(\mathbf{x},\mathbf{x}') = \left|\mathbb{E}_z[e^{\,i\Delta f_z}]\right|^2, \qquad \Delta f_z = f_z(s\mathbf{x}) - f_z(s\mathbf{x}'). \qquad (7)$$

Under the ZZ convention, $\varphi_{ij}^{ZZ}(s\mathbf{x}) = \pi^2 - \pi s(x_i + x_j) + s^2 x_i x_j$. The constant $\pi^2$ is independent of the data and therefore **cancels exactly** in $\Delta f_z$ — not merely to leading order. Writing $\mathbf{u} = \mathbf{x} - \mathbf{x}'$ we obtain

$$\Delta f_z = s\, g_z(\mathbf{u}) + s^2 h_z(\mathbf{x},\mathbf{x}'), \qquad (8)$$

with

$$g_z(\mathbf{u}) = \sum_{i\in V} u_i\, z_i - \pi \sum_{(i,j)\in E} (u_i + u_j)\, z_i z_j, \qquad (9)$$

$$h_z(\mathbf{x},\mathbf{x}') = \sum_{(i,j)\in E} (x_i x_j - x_i{}' x_j{}')\, z_i z_j. \qquad (10)$$

Both $g_z$ and $h_z$ are supported on the non-empty subsets $\{i\}$ and $\{i,j\}$, so by (5) $\mathbb{E}_z[g_z] = \mathbb{E}_z[h_z] = 0$.

**Proposition 1 (Induced metric).** *Let $G = (V,E)$ have adjacency matrix $A$ and degree matrix $\mathcal{D}$, and let $Q = \mathcal{D} + A$ be its signless Laplacian. Under the ZZ convention with $L = 1$,*

$$1 - K(\mathbf{x},\mathbf{x}') = s^2\, \mathbf{u}^\top M\, \mathbf{u} + \mathcal{O}(s^3), \qquad M = I + \pi^2 Q, \qquad (11)$$

*with $\mathbf{u} = \mathbf{x} - \mathbf{x}'$. Explicitly, $M_{ii} = 1 + \pi^2\deg(i)$ and $M_{ij} = \pi^2$ for $(i,j) \in E$, zero otherwise.*

*Proof.* Expanding (7) using (8),

$$\mathbb{E}_z[e^{i\Delta f_z}] = 1 + is\,\mathbb{E}_z[g_z] + is^2\mathbb{E}_z[h_z] - \frac{s^2}{2}\mathbb{E}_z[g_z^2] + \mathcal{O}(s^3). \qquad (12)$$

The two first-moment terms vanish, leaving an expression that is real through order $1 - \frac{s^2}{2}\mathbb{E}_z[g_z^2] + \mathcal{O}(s^3)$; taking the modulus square gives $K = 1 - s^2\,\mathbb{E}_z[g_z^2] + \mathcal{O}(s^3)$. The singletons $\{i\}$ and the edges $\{i,j\}$ are pairwise distinct subsets, so orthonormality (5) applied to (9) gives

$$\mathbb{E}_z[g_z^2] = \sum_{i\in V} u_i^2 + \pi^2 \sum_{(i,j)\in E} (u_i + u_j)^2 = \mathbf{u}^\top\mathbf{u} + \pi^2\, \mathbf{u}^\top Q\, \mathbf{u}, \qquad (13)$$

where the last step is the standard edge form of the signless Laplacian, $\mathbf{u}^\top Q\,\mathbf{u} = \sum_{(i,j)\in E}(u_i + u_j)^2$. ■

Three consequences follow immediately.

**Corollary 1 (Positive definiteness).** *The edge form $\mathbf{u}^\top Q\,\mathbf{u} = \sum_{(i,j)\in E}(u_i + u_j)^2$ is* manifestly non-negative, so $Q \succcurlyeq 0$ (Cvetković et al., 2007) and *hence $M \succcurlyeq I \succ 0$. The induced quadratic form is a genuine Mahalanobis metric for every entanglement graph, and the surrogate kernel of Section 2.1.3 is positive definite by construction.*

**Corollary 2 (Anisotropy is graph spectrum).** *The eigenvalues of $M$ are $1 + \pi^2\mu_i$, where $\mu_1 \le \cdots \le \mu_k$ are the signless Laplacian eigenvalues of $G$. The anisotropy of the induced metric is thus entirely determined by the spectral spread of the entanglement graph. Since $\mu_1 = 0$ precisely when $G$ has a bipartite connected component (Desai & Rao,*

*1994), the path $P_k$ and even cycles admit a direction along which the kernel reduces to the plain Euclidean metric.* The characterisation of when the least signless-Laplacian eigenvalue vanishes is classical (Desai & Rao, 1994); see Cvetković et al. (2007) for a systematic account of the signless Laplacian.

**Corollary 3 (Dense entanglement is least anisotropic).** *For the complete graph, $Q = (k-2)I + J$ with $J$ the all-ones matrix, so $M$ has only two distinct eigenvalues, $1 + \pi^2(k-2)$ with multiplicity $k-1$ and $1 + \pi^2(2k-2)$. Full entanglement therefore induces an almost isotropic metric perturbed by a single rank-one direction.*

Figure 1 displays the induced metric and its spectrum for the three topologies, making the content of Proposition 1 and its corollaries directly visible.

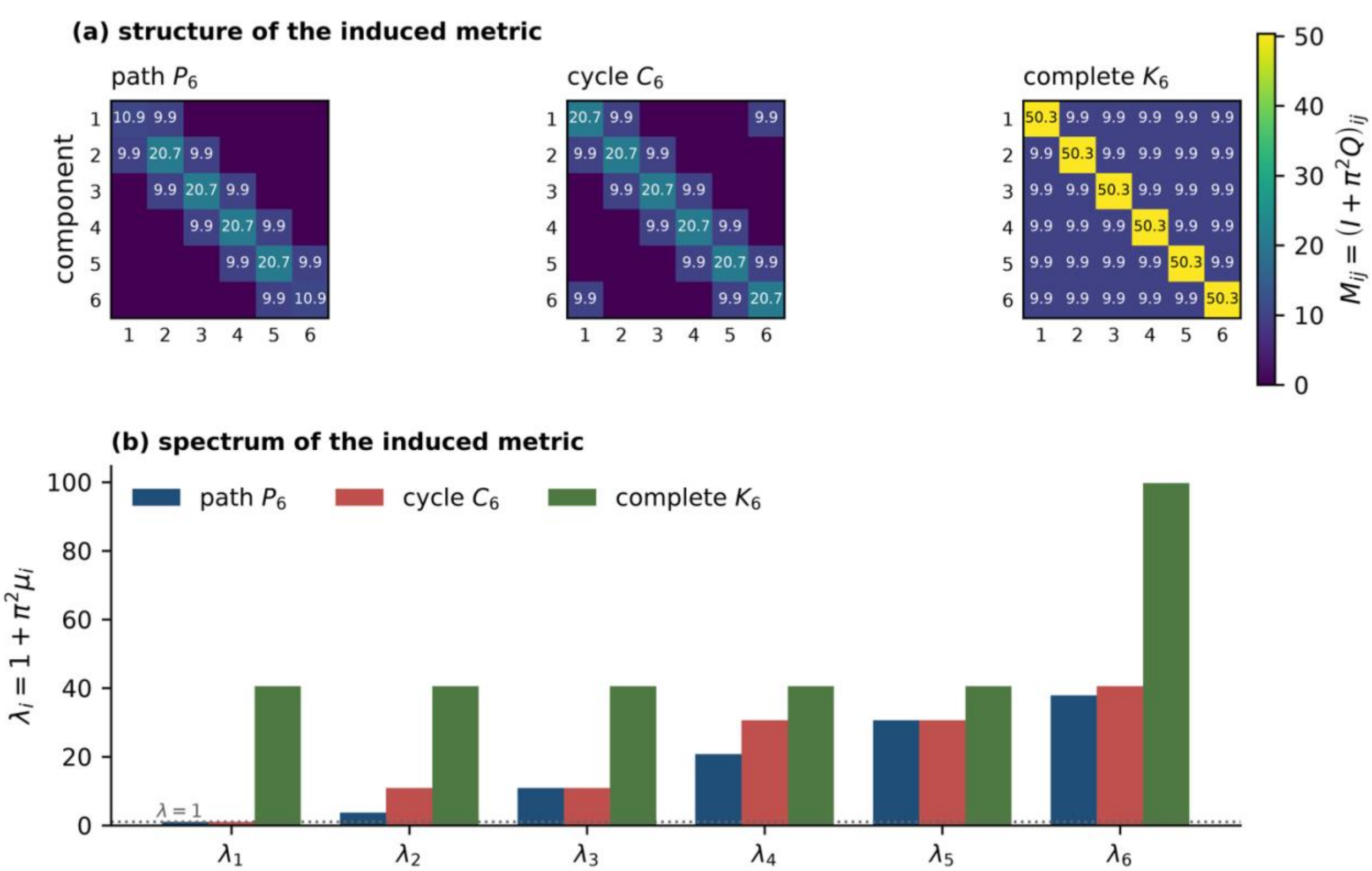

**Figure 1.** The metric induced by the ZZ feature map, $M = I + \pi^2 Q$, for three entanglement topologies at $k = 6$. **(a)** Matrix entries on a common colour scale: diagonal entries are $1 + \pi^2 \deg(i)$ , and off-diagonal entries are $\pi^2$ on the edges of $G$ and zero otherwise. **(b)** Sorted eigenvalues $\lambda_i = 1 + \pi^2 \mu_i$, with $\mu_i$ the signless Laplacian spectrum of $G$. The figure presents the content of Proposition 1 rather than data, and two consequences are visible directly. The path and the cycle are bipartite, so $\mu_1 = 0$ and each admits a direction along which the kernel reduces to the plain Euclidean metric ($\lambda_1 = 1$, dotted line; Corollary 2). The complete graph has only two distinct eigenvalues, $1 + \pi^2(k-2)$ with multiplicity $k-1$ and $1 + \pi^2(2k-2)$, so dense entanglement induces an almost isotropic metric perturbed in a single rank-one direction (Corollary 3).

**Proposition 2 (Convention dependence).** *Under the IQP convention $\varphi_{ij}^{IQP} = y_i y_j$, the pair term contributes at order $s^2$ only, so $g_z(\mathbf{u}) = \sum_i u_i z_i$ and*

$$1 - K(\mathbf{x}, \mathbf{x}') = s^2 \|\mathbf{u}\|^2 + \mathcal{O}(s^4), \qquad M = I, \qquad (14)$$

*independently of the entanglement graph. Moreover, the error term is $\mathcal{O}(s^4)$, one order smaller than in Proposition 1.*

*Proof.* Under the IQP convention $\varphi_{ij}^{IQP}(s\mathbf{x}) = s^2 x_i x_j$, so the pair phase carries no term of order $s$ and

$$\Delta f_z = s\,\gamma_z(\mathbf{u}) + s^2 h_z(\mathbf{x},\mathbf{x}'), \qquad \gamma_z(\mathbf{u}) = \sum_{i\in V} u_i\, z_i, \qquad (15)$$

with $h_z$ as in (10). Thus $\gamma_z$ is supported on the singletons $\{i\}$ and $h_z$ on the edges $\{i,j\}$, and both have zero mean by (5). Their product is supported on symmetric differences $\{i\} \triangle \{j,l\}$, which have cardinality 1 or 3 and are therefore never empty; hence

$$\mathbb{E}_z[\gamma_z h_z] = 0. \qquad (16)$$

Consequently, the $s^3$ term of $\mathbb{E}_z[e^{i\Delta f_z}]$ vanishes and

$$\mathbb{E}_z[e^{i\Delta f_z}] = 1 - \frac{s^2}{2}\mathbb{E}_z[\gamma_z^2] + \mathcal{O}(s^4), \qquad \mathbb{E}_z[\gamma_z^2] = \sum_{i\in V} u_i^2 = \|\mathbf{u}\|^2, \qquad (17)$$

by orthonormality. Squaring gives the claim.

**Remark 1 (Tightness).** *The gap between the two error orders is itself informative. In Proposition 1, the function $g_z$ carries edge terms, so the product $g_z h_z$ contains contributions $\chi_{ij}\chi_{ij} = \chi_\emptyset = 1$ with non-vanishing mean; the $\mathcal{O}(s^3)$ correction in (11) is therefore genuine and cannot be improved. Under the IQP convention, this channel is closed by (16). Numerically, the residual of a quadratic fit to $1-K$, divided by $s^2$, scales as $s$ for the ZZ convention and as $s^2$ for the IQP convention, confirming both statements.*

We highlight the contrast between Propositions 1 and 2. The anisotropy of the ZZ map arises entirely from the $\pi$-shift in (4), which promotes an entangling contribution from order $s^2$ to order $s$. Under the unshifted convention, the small-bandwidth kernel is an isotropic Gaussian on the input space, regardless of the entanglement pattern. This result provides an analytical account of the empirical observation, reported under that convention, that bandwidth-tuned quantum kernels closely resemble radial basis function kernels (Flórez-Ablan et al., 2025).

### *3. The explicit classical surrogate*

Proposition 1 suggests replacing the circuit by

$$K_{twin}(\mathbf{x},\mathbf{x}') = \exp(-s^2\,\mathbf{u}^\top M\,\mathbf{u}), \qquad M = I + \pi^2 Q, \qquad (18)$$

which agrees with $K$ to $\mathcal{O}(s^3)$ because $e^{-a} = 1 - a + \mathcal{O}(a^2)$. We stress that (18) uses a graph metric fixed by the circuit: $M$ is read off the entanglement graph, and the single free quantity $s$ is selected by the same cross-validation grid used for the quantum kernel. Evaluating $K_{twin}$ costs $\mathcal{O}(k^2)$ per pair and requires no quantum simulation.

### *4. Deeper circuits*

For $L \geq 2$ the factorisation leading to (11) no longer applies, because $D$ and $H^{\otimes k}$ do not commute and, under the ZZ convention, $D(\mathbf{0}) \neq I$. The quadratic structure nevertheless survives, for a reason independent of the particular circuit architecture.

**Proposition 3 (Persistence of the quadratic form).** *Write $f_z(s\mathbf{x}) = c_z + s\,\gamma_z(\mathbf{x}) + s^2\eta_z(\mathbf{x})$, where $\gamma_z$ is linear in $\mathbf{x}$. Then $|\phi(s\mathbf{x})\rangle = |\psi_0\rangle + s|w(\mathbf{x})\rangle + \mathcal{O}(s^2)$ with $|w(\mathbf{x})\rangle$ linear in $\mathbf{x}$, and consequently*

$$1 - K(\mathbf{x},\mathbf{x}') = s^2\,\Delta w^\dagger(\mathbb{1} - |\psi_0\rangle\langle\psi_0|)\Delta w + \mathcal{O}(s^3), \qquad (19)$$

*$\Delta w = w(\mathbf{x}) - w(\mathbf{x}')$, which is again a quadratic form in $\mathbf{u}$. Equivalently, the induced metric is the pullback of the Fubini–Study metric along the data-encoding map.*

***Proof (sketch).*** Since $f_z(s\mathbf{x}) = c_z + s\,\gamma_z(\mathbf{x}) + s^2\eta_z(\mathbf{x})$ with $\gamma_z(\mathbf{x})$ linear in $\mathbf{x}$, the diagonal gate admits the expansion $D(s\mathbf{x}) = D_0\big(I - is\Gamma(\mathbf{x}) + \mathcal{O}(s^2)\big)$, where $\Gamma(\mathbf{x}) = \mathrm{diag}\big(\gamma_z(\mathbf{x})\big)$ is linear in $\mathbf{x}$. Writing a single layer as $A(s\mathbf{x}) =$ $D(s\mathbf{x})H^{\otimes k} = A_0 + sB(\mathbf{x}) + O(s^2)$, the first-order correction to the L-layer circuit is $V_L(x) = \sum_{\ell=1}^{L} {A_0}^{L-\ell} B(x) {A_0}^{\ell-1}$, which is a sum over the L possible layer positions at which $\Gamma(\mathbf{x})$ is inserted, each surrounded by x-independent operators. Consequently, $V_L(\mathbf{x})$ — and hence $|w(\mathbf{x})\rangle = V_L(\mathbf{x})\,|0^{\otimes k}\rangle$ — is linear in $\mathbf{x}$.

Unitarity of the circuit forces $\mathrm{Re}\langle\psi_0|w(\mathbf{x})\rangle = 0$, so that the second-order state corrections cancel in $1 - \mathrm{K}$ and the surviving term is $s^2\|P(|\,w(\mathbf{x})\rangle - |\,w(\mathbf{x}')\rangle)\|^2$ with $P = I - |\,\psi_0\rangle\langle\,\psi_0|$. Since $|\,w(\mathbf{x})\rangle$ is linear in $\mathbf{x}$, we have $|\,w(\mathbf{x})\rangle - |\,w(\mathbf{x}')\rangle = |\,w(\mathbf{u})\rangle$, yielding $1 - K = s^2\, u^\top M_L\, u + O(s^3)$, where the fixed matrix $(M_L)_{ij} = \mathrm{Re}\langle w(e_i)|P|w(e_j)\rangle$ is the Fubini–Study pullback evaluated at y = 0 in the unscaled coordinate. The full proof is deferred to Appendix A.

Proposition 3 guarantees that a classical Mahalanobis surrogate exists for every depth, with at most $\frac{k(k+1)}{2}$ parameters, but it does not supply $M$ in closed form. For $L \geq 2$, the Hadamard layers propagate correlations beyond the edges of $G$ and the resulting metric is no longer a function of $Q$ alone; we determine it numerically in Section 3 and leave a closed form for future work.

### B. Datasets and preprocessing

We use two near-infrared spectroscopic regression benchmarks. Tecator comprises 215 finely chopped meat samples measured over 100 wavelength channels in the range 850–1050 nm, with three response variables (fat, water, and protein content), each expressed as a percentage. Gasoline comprises 60 samples measured over 401 channels from 900 to 1700 nm at 2 nm intervals, with octane number as the response. Both are standard benchmarks in chemometrics and are distributed with the R packages fda.usc (Febrero-Bande & Oviedo de la Fuente, 2012) and pls (Mevik & Wehrens, 2007), respectively; the Tecator data were originally described by Borggaard & Thodberg (1992) and the Gasoline data by Kalivas (1997). Together they span a useful range of regimes: Tecator has more samples than channels and a well-known nonlinear response for the fat target, whereas Gasoline is strongly underdetermined with roughly seven channels per sample.

Spectral preprocessing is treated as a hyperparameter rather than fixed in advance. We consider five pipelines: the raw absorbances (raw); standard normal variate correction (snv), which centres and scales each spectrum individually to remove multiplicative scatter; first and second Savitzky–Golay derivatives with a window of 11 channels and a second-order polynomial (sg1, sg2); and standard normal variate correction followed by the second derivative (sg2s). Each of these operates either within a spectrum or along the wavelength axis, so none of them uses response values or information across samples, and applying them before splitting introduces no leakage.

The decision to treat preprocessing as a tunable choice is not cosmetic. Within a single target, different kernel families prefer different pipelines: for the protein target, the Matérn, isotropic RBF, and ARD-RBF families select the second-derivative pipeline in 80–94% of splits, while the quantum kernel and its twin select the raw spectra in 46–47%. Fixing one pipeline for all families would therefore handicap some and not others, and the direction of the resulting bias is not predictable in advance.

After preprocessing, each split is processed as follows. A standardiser is fitted on the training spectra and applied to both partitions, so that each channel has zero mean and unit variance with respect to the training set. Principal component analysis is then fitted on the standardised training spectra and both partitions are projected onto the leading $k$ components, with $k = 6$ throughout unless stated otherwise; because the encoding uses one qubit per feature, $k$ is simultaneously the number of retained components and the number of qubits. Finally, each component is divided by its maximum absolute value on the training set and the result is clipped to $[-1,1]$, which places the data in the domain the feature map expects. All fitting is confined to the training partition.

### C. Kernel families and hyperparameter grids

Six families compete on equal terms. Each receives exactly 36 kernel hyperparameter settings, crossed with the five preprocessing pipelines for a total of 180 candidates per family.

- **Quantum.** The ZZ feature map of Section 2.1 with $L = 1$ and linear entanglement, evaluated at 36 bandwidths spaced logarithmically over $[10^{-4}, 1]$. Kernels are computed from exact state vectors.
- **Twin.** The explicit surrogate $\exp(-s^2\mathbf{u}^\top M\mathbf{u})$ with $M = I + \pi^2 Q$ read off the same entanglement graph, evaluated on the *same* 36 bandwidths.

- **PQK.** A projected quantum kernel $\exp(-\gamma\, d^2)$ on the single-qubit Bloch vectors of the encoded states, with six bandwidths drawn from the quantum grid crossed with six values of $\gamma$; the $\gamma$ grid is divided by the median squared Bloch distance at each bandwidth, so that the effective scale does not drift with qubit count.
- **Isotropic RBF** and **Matérn 5/2** (Rasmussen & Williams, 2006, §4.2), each over 36 length scales.
- **ARD-RBF**, an anisotropic Gaussian kernel in the spirit of automatic relevance determination (Rasmussen & Williams, 2006, §5.1), using per-component length scales $\ell_d = c\,\sigma_d^a$, where $\sigma_d$ is the training standard deviation of component $d$; six values of $c$ crossed with six values of $a \in [0,1]$, so that $a = 0$ recovers an isotropic kernel and $a = 1$ sets each length scale proportional to the component's standard deviation, giving each component a weight in the metric that is inversely proportional to its variance. Rather than learning $k$ independent length scales, we tie them to the component standard deviations through two parameters, which keeps the candidate count equal to that of the other families.

The classical length-scale grids are anchored to the data rather than fixed. Writing $\bar{d}$ for the median pairwise distance among the training scores, both the isotropic grid and the ARD scale parameter $c$ span $\bar{d} \cdot 10^{[-1.5,\,1.5]}$. This choice was not the initial one, and the reason for changing it is worth recording: with a conventional fixed grid, every classical baseline selected the largest available length scale in every split of a preliminary run, and widening the grid changed which family attained the lowest test error. Anchoring to $\bar{d}$ removes that boundary effect without introducing a tuned constant.

Four further methods are reported as reference points rather than competitors, and are marked in the tables. These are ridge regression with a linear kernel on the principal component scores and on the full standardised spectra; polynomial kernels of degree two and three on the scores; and partial least squares on the full spectra, the de facto standard in chemometrics, with the number of latent variables selected by cross-validation over $1, \ldots, 60$. Their candidate sets are necessarily smaller because they have few or no kernel hyperparameters, and they are excluded from any comparison that depends on equal candidate counts.

**D. Estimation and evaluation protocol**

For each kernel, we solve kernel ridge regression over 40 regularisation parameters spaced logarithmically over $[10^{-10}, 10^{2}]$. Writing $K = Q\Lambda Q^{\top}$ for the eigendecomposition of the training Gram matrix, all 40 solutions and their effective degrees of freedom $\sum_d \lambda_d / (\lambda_d + \alpha)$ follow from a single decomposition, which makes the full grid inexpensive. Kernels whose off-diagonal entries have standard deviation below $10^{-12}$ are treated as degenerate and skipped.

The regularisation parameter and the kernel setting are selected jointly by five-fold cross-validation on the training partition, with fold assignment fixed across families and settings so that all candidates are scored on identical folds. Because cross-validation scores computed this way are comparable across preprocessing pipelines (the response, the folds, and the split are identical, and only the design matrix changes), selecting over the pooled set of (pipeline, setting) pairs is equivalent to selecting the best setting within each pipeline and then taking the minimum across pipelines, and we implement the latter.

Each configuration is evaluated over 100 random 70/30 train–test splits, with the split seed determined by the replicate index so that every family and every preprocessing pipeline sees exactly the same partitions. Reported test error is the mean squared error at the cross-validated optimum.

Comparisons between families are paired on the split index and summarised by the mean difference with a 95% percentile bootstrap interval over 20,000 resamples of the split-level differences, together with the fraction of splits on which each family wins. We deliberately do not report rank tests of the difference. The replicates are repeated partitions of one fixed dataset rather than independent draws from a population, so a $p$-value computed across them has no sampling interpretation, and its magnitude is governed by the number of resampling replicates we chose to run. For the same reason, we characterise agreement between the twin and the quantum kernel by effect size (the median $|\log ratio|$ of test errors) rather than by whether an interval excludes zero.

### E. Diagnosing the classical reduction

To measure how completely a simulated kernel is described by a quadratic form, we work with the kernel distance. Since $K_{ii} = 1$ for a fidelity kernel, $d_K^2(i,j) = 2\left(1 - K_{ij}\right)$, and Proposition 1 predicts that this equals $2s^2\mathbf{u}^\top M\mathbf{u}$ to leading order. We therefore regress $d_K^2$ over all training pairs on three nested designs in the coordinate differences $\Delta z$:

1. **Isotropic**, a single predictor $\sum_d \Delta z_d^2$;
2. **Diagonal**, the $k$ predictors $\Delta z_d^2$, which is the ARD form;
3. **Full quadratic**, adding the $k(k-1)/2$ interaction terms $\Delta z_d \, \Delta z_e$.

The coefficient of determination of each design measures how much of the kernel geometry that class of classical metric can express, and the increments between them separate the contribution of anisotropy from that of the pairwise interactions the entangling gates produce. Fitting uses ordinary least squares, since interaction coefficients may take either sign. With $k = 6$ , the full design has 21 parameters against several thousand pairs, so the fits are not at risk of interpolation.

Where a fitted metric is compared with the analytical prediction, we report the maximum absolute entry difference and the same quantity relative to the largest entry of $M$. Where the recovered length-scale profile is compared with the one ARD-RBF selects, we use Spearman correlation on the logarithms, which is undefined and reported as missing when ARD selects $a = 0$ , and its profile is constant.

### F. Computational implementation and convention verification

Because the central result of Section 2.1 depends on a specific phase convention, we state the implementation precisely.

All quantum kernels are computed from exact state vectors using Qiskit 2.5.2. The feature map is constructed directly rather than through the library class, so that the circuit is fully specified here: each layer applies a Hadamard to every qubit, then $R_Z(2sx_i)$ on qubit $i$, and then, for each edge $(i,j) \in E$ of the entanglement graph, the block $\mathrm{CX}(i,j)\, R_Z\left(2\varphi_{ij}(s\mathbf{x})\right)\mathrm{CX}(i,j)$, with $\varphi_{ij}$ as in equation (4). We use the convention $R_Z(\theta) = \exp(-i\theta Z/2)$, so that $R_Z(2y_i)$ implements $\exp(-iy_i Z_i)$ and the CX–$R_Z$–CX block implements $\exp\left(-i\varphi_{ij} Z_i Z_j\right)$; this is what makes equation (2) the exact diagonal of the layer. Global phases are immaterial because the kernel is a squared overlap.

Two consequences of this construction are worth making explicit. First, the $\pi$-shifted pair phase $\varphi_{ij}^{\mathrm{ZZ}}(\mathbf{y}) = (\pi - y_i)\left(\pi - y_j\right)$ is the default of the ZZFeatureMap class in Qiskit; Proposition 1 applies to that convention, and Proposition 2 to the unshifted one. Second, because the constant $\pi^2$ term of $\varphi_{ij}^{\mathrm{ZZ}}$ is data-independent, the diagonal $D(\mathbf{0})$ is not the identity; this is why the single-layer factorisation of Section 2.1.2 does not extend directly to $L \geq 2$.

The correspondence between the simulated circuit and the analytical expansion is verified numerically rather than assumed. Two checks in the deposited code establish it: one recovers the fitted quadratic form from the simulated kernel for the path, cycle and complete entanglement graphs and compares it entrywise with $I + \pi^2 Q$ (Table I), and one repeats the fit across circuit depths $L = 1,2,3$ under both phase conventions (Table II), where the unshifted convention returns off-diagonal entries that vanish identically. A convention error in either direction would be visible immediately in these tables: under the wrong phase convention the recovered metric would be the identity rather than $I + \pi^2 Q$, and under the wrong $R_Z$ normalisation the recovered entries would be off by a factor of two.

## III. Results

We report the results as follows. Section 3.1 verifies the closed-form metric of Proposition 1 against direct simulation. Section 3.2 examines circuit depth and the phase convention. Section 3.3 locates the boundary of the classical regime. Section 3.4 tests the predictive equivalence of the classical twin, Section 3.5 shows that the non-classical regime adds no robust predictive value, and Section 3.5.1 traces its one exception to the retained dimension. Section 3.6 places the quantum kernel among conventional classical baselines.

## A. Numerical verification of the closed-form metric

Table 1 compares the metric predicted by Proposition 1 against the quadratic form recovered by least squares from the simulated kernel, for three entanglement topologies at $s = 0.002$ and $k = 6$ qubits. In every case, the full quadratic model (diagonal terms plus all pairwise cross terms) reproduces $1 - K$ essentially exactly, and the recovered matrix agrees with $I + \pi^2 Q$ to within a relative error below $10^{-3}$.

***Table I.*** *Verification of* $M = I + \pi^2 Q \quad (k = 6, s = 0.002, N = 90\ points)$.

| Topology | $\|E\|$ | $R^2$ (quadratic) | $\max\|M_{\text{fit}} - M\|$ | relative |
|---|---|---|---|---|
| Path $P_6$ | 5 | 1.000000 | $7.75 \times 10^{-3}$ | $3.7 \times 10^{-4}$ |
| Cycle $C_6$ | 6 | 1.000000 | $9.31 \times 10^{-3}$ | $4.5 \times 10^{-4}$ |
| Complete $K_6$ | 15 | 0.999999 | $4.46 \times 10^{-2}$ | $8.9 \times 10^{-4}$ |

The recovered matrices reproduce the predicted structure entry by entry. For the path graph, the fit returns $M_{11} = M_{66} = 10.87$, $M_{ii} = 20.74$ for interior vertices and $M_{i,i+1} = 9.87$, matching $1 + \pi^2$, $1 + 2\pi^2$ and $\pi^2$; all entries at graph distance greater than one are zero to fitting precision. For the cycle, the corner entries $M_{16} = M_{61} = 9.87$ appear as the chain closes, and every diagonal entry equals $1 + 2\pi^2$ because no vertex is now an endpoint. For the complete graph, all diagonal entries equal $1 + 5\pi^2 = 50.35$ and every off-diagonal entry equals $\pi^2$.

Table I summarises the agreement; Figure 2 shows it pair by pair.

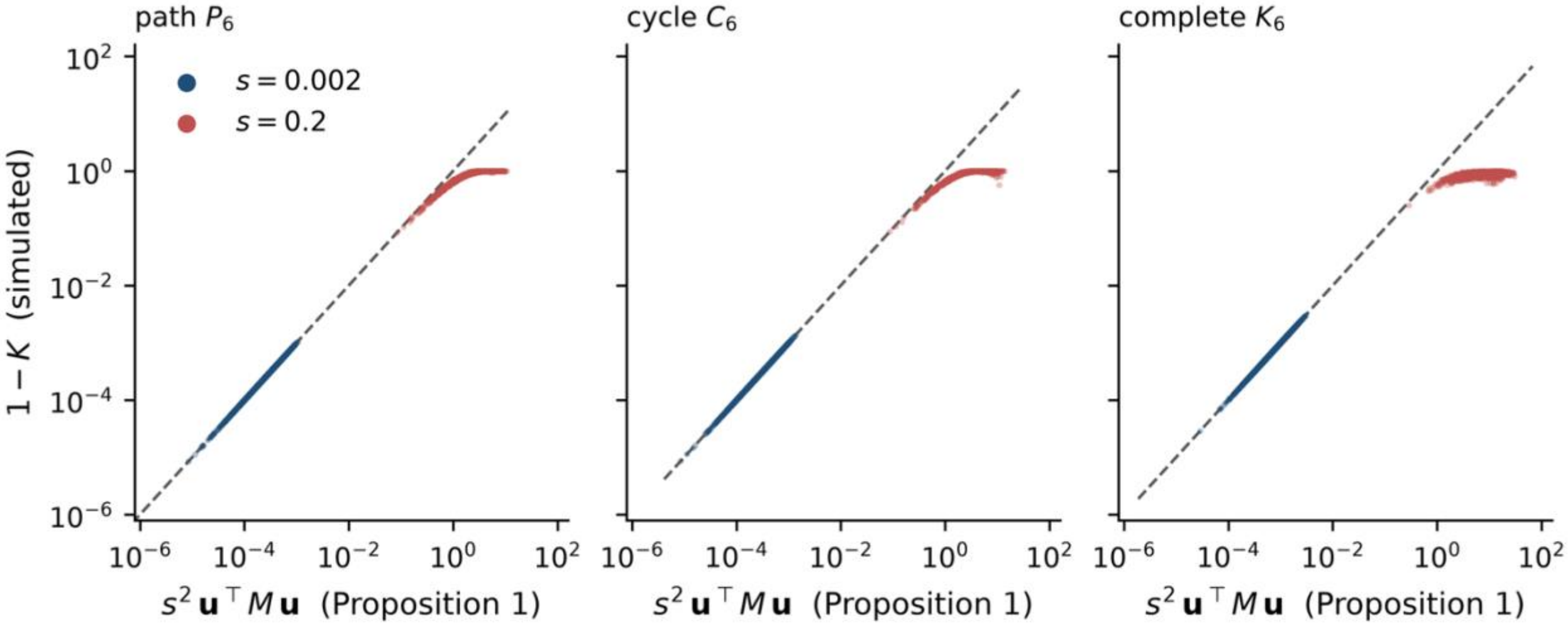

**Figure 2.** Direct verification of Proposition 1. Each point is one training pair $(\mathbf{x}, \mathbf{x}')$: the horizontal axis is the closed-form prediction $s^2 \mathbf{u}^\top M \mathbf{u}$, with $M$ read off the entanglement graph and no quantity fitted to the simulation, and the vertical axis is $1 - K$ obtained from exact state-vector simulation. Both axes are logarithmic, and the dashed line is the identity. At $s = 0.002$, inside the classical regime, the points follow the identity over more than two decades for all three topologies. At $s = 0.2$, beyond the boundary located in Section 3.3, the simulated values saturate and depart from the prediction. Here $k = 6$ with 70 points, giving 2415 pairs per cloud.

## B. Depth and phase convention

Table II reports the same fit for one, two, and three circuit layers under both phase conventions. Three observations follow.

*Table II. Quadratic reduction across depth and convention* ($k = 6, s = 0.002$).

| Convention | $L$ | $R^2$ isotropic | $R^2$ quadratic | $1 - R^2$ | off-diagonal / diagonal |
|---|---|---|---|---|---|
| ZZ | 1 | 0.5078 | 1.000000 | $2.1 \times 10^{-7}$ | 0.476 |
| ZZ | 2 | 0.4682 | 0.999978 | $2.3 \times 10^{-5}$ | 0.532 |
| ZZ | 3 | 0.5362 | 0.999978 | $2.2 \times 10^{-5}$ | 0.593 |
| IQP | 1 | 1.000000 | 1.000000 | $7.9 \times 10^{-11}$ | 0.000 |
| IQP | 2 | 1.000000 | 1.000000 | $1.8 \times 10^{-10}$ | 0.000 |
| IQP | 3 | 1.000000 | 1.000000 | $1.2 \times 10^{-9}$ | 0.000 |

First, the quadratic reduction is not an artefact of the single-layer case: under the ZZ convention $R^2$ remains above $0.99997$ at $L = 3$, as Proposition 3 requires. Second, an isotropic model is badly misspecified for the ZZ convention at every depth ($R^2 \approx 0.5$), so the reduction genuinely requires the anisotropic metric. Third, and confirming Proposition 2, the IQP convention yields an isotropic $R^2$ of exactly one and off-diagonal entries that vanish identically, at all three depths and independently of the entanglement pattern.

The effect of depth is to widen the support of the metric. Under the ZZ convention, the mean absolute band profile $|M_{i,i+d}|$ is $(20.74, 9.87, 0, 0, 0)$ at $L = 1$, $(28.17, 16.55, 2.98, 3.26, 0.03)$ at $L = 2$ and $(47.50, 22.72, 4.11, 7.54, 6.66)$ at $L = 3$ for $d = 0, \ldots, 4$. The interleaved Hadamard layers propagate correlations beyond the edges of $G$, so that circuit depth acts as a control on the interaction range of the induced classical metric.

**C. The boundary of the classical regime**

Figure 3(a) shows the median $R^2$ of the quadratic fit as a function of the bandwidth, computed over all splits and preprocessing pipelines for each dataset, and Table III lists representative values. The reduction is essentially exact ($R^2 \geq 0.999999$) for $s \lesssim 7 \times 10^{-3}$, degrades slowly through $R^2 = 0.998$ at $s = 0.055$, and collapses beyond $s \approx 0.16$.

The crossing of $R^2 = 0.99$ occurs at exactly the same grid point for both datasets: the last bandwidth with $R^2 \geq 0.99$ is $s = 0.0720$ and the first below it is $s = 0.0936$. Since the two datasets differ in sample size, dimension, and target, this coincidence is consistent with the boundary being a property of the encoding circuit rather than of the data. The leading deviation is of order $s^3$, with a prefactor that depends on the circuit and on the bounded input domain. We use $s^* = 0.072$ as the nominal boundary in what follows, and Section 3.5 confirms that no conclusion depends on this choice.

Panels (b) and (c) of Figure 3 anticipate the two results that follow. The predictive disagreement between the twin and the quantum kernel, plotted on the same bandwidth axis, rises in step with the departure of $R^2$ from unity, so that the two curves locate the same transition. The selection histograms show that cross-validation concentrates well below $s^*$ for the fat and octane targets, is bimodal for water, and is nearly flat for protein. We discuss this difference in Section 3.5.1.

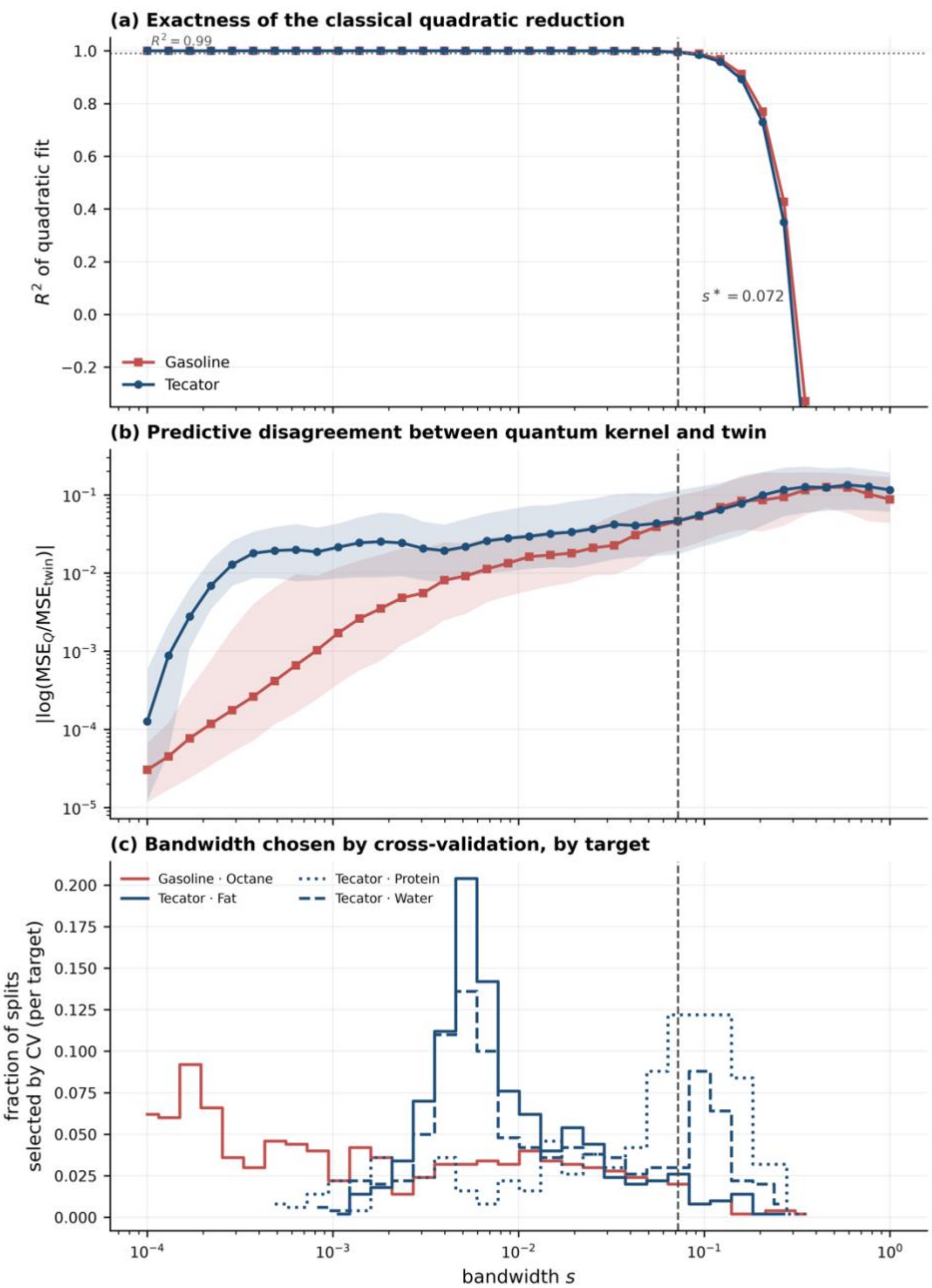

**Figure 3.** Boundary of the classical regime and its consequences, as a function of the bandwidth $s$ (logarithmic axis; dashed vertical line at $s^* = 0.072$). **(a)** Median $R^2$ of the full quadratic fit to $1 - K$; the dotted line marks $R^2 = 0.99$. **(b)** Median $\left|\log\left(\mathrm{MSE}_Q/\mathrm{MSE}_{\mathrm{twin}}\right)\right|$ evaluated at the same bandwidths, with interquartile bands. **(c)** Distribution of the bandwidth selected by cross-validation, shown separately for each target. Panels (a) and (b) rise together: the twin and the quantum kernel begin to disagree exactly where the quadratic reduction ceases to be exact.

***Table III.*** *Median $R^2$ of the quadratic fit by bandwidth.*

| $s$ | Gasoline | Tecator |
|---|---|---|
| $1.0 \times 10^{-4}$ | 1.000000 | 1.000000 |
| $8.2 \times 10^{-4}$ | 1.000000 | 1.000000 |
| $6.7 \times 10^{-3}$ | 1.000000 | 0.999999 |
| $1.9 \times 10^{-2}$ | 0.999979 | 0.999967 |
| $5.5 \times 10^{-2}$ | 0.998653 | 0.998035 |
| $1.6 \times 10^{-1}$ | 0.912390 | 0.892661 |
| $4.5 \times 10^{-1}$ | −1.977 | −2.547 |

**D. The classical twin matches the quantum kernel**

Table IV compares the explicit twin of equation (18) against the quantum kernel across two datasets, four targets, and five preprocessing pipelines, with 100 resampled splits in each of the 20 cells. Both kernels select their bandwidth and ridge parameter by the same cross-validation protocol.

***Table IV.*** *Twin versus quantum kernel: mean test MSE, paired difference (quantum − twin), 95% percentile bootstrap interval, quantum win rate and mean $|log\ ratio|$.*

| Dataset | Target | Mode | Quantum | Twin | Diff. | 95% percentile-bootstrap interval | Win | ⟨\|log\|⟩ |
|---|---|---|---|---|---|---|---|---|
| Gasoline | Octane | raw | 0.0973 | 0.0977 | −0.0004 | [−0.0011, +0.0004] | 0.46 | 0.020 |
| | | snv | 0.1322 | 0.1323 | −0.0001 | [−0.0020, +0.0022] | 0.52 | 0.026 |
| | | sg1 | 0.1205 | 0.1218 | −0.0013 | [−0.0067, +0.0038] | 0.56 | 0.063 |
| | | sg2 | 0.0865 | 0.0845 | +0.0019 | [−0.0010, +0.0055] | 0.48 | 0.058 |
| | | sg2s | 0.1072 | 0.1075 | −0.0002 | [−0.0055, +0.0042] | 0.52 | 0.060 |
| Tecator | Fat | raw | 1.8112 | 1.8662 | −0.0550 | [−0.2011, +0.0738] | 0.38 | 0.126 |
| | | snv | 1.1458 | 1.1958 | −0.0500 | [−0.1688, +0.0304] | 0.47 | 0.088 |
| | | sg1 | 1.6759 | 1.5473 | +0.1286 | [+0.0448, +0.2241] | 0.40 | 0.114 |
| | | sg2 | 2.3944 | 2.2418 | +0.1526 | [−0.0406, +0.3903] | 0.42 | 0.142 |
| | | sg2s | 1.3149 | 1.2159 | +0.0990 | [−0.0064, +0.2307] | 0.49 | 0.110 |
| Tecator | Water | raw | 2.8379 | 2.8409 | −0.0030 | [−0.3053, +0.2257] | 0.33 | 0.120 |
| | | snv | 2.7170 | 2.7496 | −0.0327 | [−0.1780, +0.1091] | 0.44 | 0.116 |
| | | sg1 | 2.8384 | 2.8521 | −0.0138 | [−0.0977, +0.0825] | 0.59 | 0.086 |
| | | sg2 | 3.4493 | 3.3511 | +0.0982 | [−0.0176, +0.2218] | 0.47 | 0.090 |
| | | sg2s | 2.7540 | 2.7507 | +0.0033 | [−0.0916, +0.0911] | 0.42 | 0.100 |
| Tecator | Protein | raw | 1.0409 | 1.0268 | +0.0141 | [−0.0029, +0.0365] | 0.35 | 0.031 |
| | | snv | 1.2908 | 1.3109 | −0.0200 | [−0.0684, +0.0230] | 0.56 | 0.073 |
| | | sg1 | 1.2976 | 1.2833 | +0.0143 | [−0.0206, +0.0456] | 0.39 | 0.046 |
| | | sg2 | 1.2805 | 1.3044 | −0.0239 | [−0.0658, +0.0062] | 0.45 | 0.044 |
| | | sg2s | 1.1101 | 1.1770 | −0.0668 | [−0.1640, −0.0009] | 0.44 | 0.076 |

The bootstrap interval contains zero in 18 of the 20 cells. The two exceptions (Tecator fat under first-derivative preprocessing, where the twin is better, and Tecator protein under SNV followed by second-derivative preprocessing, where the quantum kernel is better) lie in opposite directions, which is consistent with sampling variability rather than a systematic gap. Three further summaries support that reading: the mean of the cell-level log ratios is 9 positive and $11$ negative, the pooled mean log ratio is $+0.0033$, and the pooled quantum win rate is $0.457$.

We therefore characterise the agreement by effect size rather than by significance. The median $|\log(\mathrm{MSE}_Q/\mathrm{MSE}_{\mathrm{twin}})|$ over all 2000 splits is $0.0265$: a typical split produces test errors differing by under 3%.

The residual disagreement behaves as the theory predicts. Within each dataset, $|\log ratio|$ increases with the cross-validated bandwidth, with Spearman correlations of $+0.680$ ($p = 3 \times 10^{-69}$) on Gasoline and $+0.226$ ($p = 9 \times 10^{-19}$) on Tecator. Grouping all splits by the boundary of Section 3.3 makes the same point directly: for $s \le s^*$ the median $|\log \text{ratio}|$ is $0.0227$ with a win rate of $0.473$, whereas for $s > s^*$ it rises to $0.0591$ with a win rate of $0.371$. The twin and the quantum kernel diverge exactly where Proposition 1 says they should.

**E. The non-classical regime adds no robust predictive value**

If the reduction fails only for $s > s^*$, a natural question is whether that region is nevertheless useful — whether cross-validation gains anything by being allowed to select it. It does not, with the partial exception of one target discussed below.

Table V reports the paired difference in test error between selecting the bandwidth on the full grid and selecting it on a grid truncated at a candidate threshold, over all 2000 splits. A positive difference means that truncation *improves* test error.

***Table V.*** *Cost of forbidding the non-classical regime (2000 splits; mean full-grid MSE=1.4751).*

| Threshold | Diff. | 95% percentile-bootstrap interval | Relative | Cells selecting above | Cells improved |
|---|---|---|---|---|---|
| $s \le 0.020$ | $+0.0719$ | $[+0.0434, +0.1031]$ | 4.87% | 34.2% | 19.0% |
| $s \le 0.036$ | $+0.0658$ | $[+0.0383, +0.0962]$ | 4.46% | 27.5% | 15.2% |
| $s \le 0.052$ | $+0.0620$ | $[+0.0351, +0.0921]$ | 4.21% | 24.6% | 13.9% |
| $s \le 0.072$ | $+0.0557$ | $[+0.0329, +0.0826]$ | 3.77% | 15.7% | 9.9% |
| $s \le 0.100$ | $+0.0439$ | $[+0.0231, +0.0694]$ | 2.98% | 10.0% | 6.8% |
| $s \le 0.140$ | $+0.0341$ | $[+0.0150, +0.0586]$ | 2.31% | 4.9% | 3.6% |

Apart from the Tecator protein exceptions discussed below, truncation does not degrade test error, and at every threshold it improves it by an amount whose bootstrap interval excludes zero. The effect is monotone in the threshold and holds in 16 of the 20 cells, with two cells unaffected because the upper regime is never selected there.

The mechanism is concentrated rather than diffuse. Cross-validation selects $s > s^*$ in 15.7% of splits; in exactly those splits, truncation reduces mean test error from 2.1299 to 1.7741, an improvement of 16.7%. In the remaining 84.3%, the two procedures coincide by construction. The pooled 3.77% is therefore a dilution of a large but infrequent selection failure.

A control experiment excludes the obvious alternative explanation, that the gain merely reflects a smaller selection problem. Truncating at $s^*$ reduces the grid from 36 to 26 bandwidths. Removing an equally large random subset of 26 bandwidths, repeated 300 times, changes mean test error by only $+0.0037$ (0.25%), with a 5–95% range of $[-0.0155, +0.0200]$ that comfortably covers zero; the observed effect lies above every one of the 300 random draws. Removing the ten *smallest* bandwidths instead changes test error by exactly zero. The gain is thus attributable specifically to the removal of the non-classical region and not to reduced selection overfitting.

One cell group runs against this pattern. For the Tecator protein target, cross-validation selects the upper regime in 36.2% of splits and truncation *worsens* mean test error by $-0.0993$ under SNV and $-0.0295$ under SNV with second derivative; these are the only two negative cells in the table. The magnitudes are small (about 1% of that target's mean test error) but the direction is consistent, and we trace it to its cause in Section 3.5.1.

**1. The protein exception is a property of the representation**

The protein target is the one place where truncating the grid does not help, and it is also where cross-validation selects the upper regime most often. Both facts follow from a single cause: with six principal components, that target is largely unlearnable, so the cross-validation objective is nearly flat in the bandwidth and the choice is settled by noise rather than by fit.

We test this directly by varying the number of retained components. Because the number of qubits equals the number of components in our encoding, a twenty-component circuit would require a $2^{20}$-dimensional state vector and is not simulable; we therefore use the derived twin, whose cost is $O(k^2)$ per pair. Section 3.4 licenses this substitution, and we verify it below. Table VI reports 100 resampled splits at each dimension, pooled over all five preprocessing pipelines.

***Table VI.*** *Effect of the retained dimension k. Ridge rows use ridge regression alone on the first k components; the remaining rows use the twin kernel.*

| Quantity | Target | $k=4$ | $k=6$ | $k=8$ | $k=10$ | $k=12$ | $k=16$ | $k=20$ | full |
|---|---|---|---|---|---|---|---|---|---|
| Ridge test MSE | Fat | 10.72 | 8.22 | 7.84 | 7.22 | 6.87 | 6.42 | 6.12 | 6.60 |
| | Water | 7.66 | 6.43 | 6.04 | 5.70 | 5.43 | 5.13 | 5.13 | 5.50 |
| | Protein | 2.08 | 1.58 | 1.31 | 1.13 | 0.83 | 0.51 | **0.45** | 0.47 |
| CV flatness | Fat | 11.91 | 43.47 | 49.03 | 36.48 | 32.10 | 28.30 | 23.05 | — |
| | Water | 4.04 | 14.21 | 21.38 | 23.56 | 23.36 | 22.98 | 18.27 | — |
| | Protein | 0.79 | **2.71** | 3.77 | 6.63 | 9.58 | 14.40 | **15.27** | — |
| Upper/lower CV ratio | Fat | 1.048 | 1.659 | 1.681 | 1.256 | 1.388 | 1.787 | 2.064 | — |
| | Water | 1.014 | 1.150 | 1.049 | 1.111 | 1.262 | 1.543 | 1.731 | — |
| | Protein | 0.987 | **1.013** | 1.058 | 1.091 | 1.172 | 1.778 | **2.359** | — |
| Upper regime selected (%) | Fat | 37.0 | 5.4 | 7.0 | 13.6 | 5.8 | 1.2 | 0.0 | — |
| | Water | 46.2 | 23.6 | 37.0 | 24.2 | 9.2 | 4.8 | 1.6 | — |
| | Protein | 59.4 | **36.2** | 22.4 | 19.0 | 8.0 | 0.2 | **0.0** | — |
| Twin test MSE | Fat | 3.46 | 1.61 | 2.11 | 3.40 | 3.67 | 4.56 | 5.39 | — |
| | Water | 3.80 | 2.91 | 2.84 | 2.94 | 2.95 | 3.18 | 3.83 | — |
| | Protein | 1.74 | 1.22 | 1.12 | 0.87 | 0.65 | 0.46 | 0.43 | — |

CV flatness is $(\text{maxCV} - \text{minCV})/\text{minCV}$ over the bandwidth grid; the upper/lower ratio compares the best cross-validation score attainable above $s^*$ with the best attainable below it.

Four quantities move together for protein and for protein alone. Ridge alone improves from 1.58 at six components to 0.45 at twenty, matching what the full spectrum attains (0.47); the corresponding improvement for fat and water is 20–26%, and both remain far from their full-spectrum values. The cross-validation objective sharpens by a factor of 5.6, from a flatness of 2.71 to 15.27. The two arms of that objective separate, from a ratio of 1.013 (indistinguishable) to 2.359. And selection of the upper regime falls from 36.2% to zero. The same monotone pattern appears in all five preprocessing pipelines separately, falling from 96%, 61%, 25%, 39% and 76% at four components to at most 1% at sixteen.

The exception is therefore not evidence that the non-classical regime has value. It is evidence that a target which the representation cannot express produces a degenerate selection problem, in which the upper regime is chosen because nothing distinguishes it rather than because it helps. A weaker form of the same effect is visible for the other targets at four components, where fat and water select the upper regime in 37.0% and 46.2% of splits and fall to 0.0% and 1.6% at twenty; selection of the non-classical regime is a symptom of an inadequate representation generally, not a peculiarity of one target.

Increasing the dimension is not, however, a remedy. Twin test error for fat rises monotonically beyond six components, from 1.61 to 5.39, and water is flat then worsening, consistent with the kernel concentration expected at higher encoding dimension. Six components are near-optimal for fat and water and inadequate for protein; no single choice serves all three, and we retain $k = 6$ as the configuration under study rather than as a recommendation.

Finally, we verify the substitution of the twin for the quantum kernel in this experiment. At $k \leq 8$, where circuit simulation remains feasible, the two families agree on the upper/lower ratio to within 0.4% in median and on the binary regime choice in 90.6% of splits; the twin is systematically about 8% flatter in the flatness statistic. The direction and magnitude of every trend in Table VI is therefore reproduced by the quantum kernel wherever it can be computed.

**F. Comparison with classical baselines**

Section 3.4 establishes that the twin reproduces the quantum kernel; this section places both among conventional classical alternatives. All competing families (quantum, twin, PQK, ARD-RBF, isotropic RBF and Matérn 5/2) receive exactly 180 candidate settings (36 kernel hyperparameter values crossed with five preprocessing pipelines), and the preprocessing pipeline is itself selected by cross-validation rather than fixed in advance. This matters: within a single target, different families select different pipelines. For the protein target, for example, the Matérn, isotropic RBF and ARD-RBF families select second-derivative preprocessing in 80–94% of splits, while the quantum kernel and its twin select the raw spectra in 46–47%; fixing a single pipeline would therefore have handicapped some families and not others. Four reference baselines are reported for context rather than as competitors and are marked with a bullet; their smaller candidate sets (5, 10 and 300 settings) reflect the fact that they have few or no kernel hyperparameters.

***Table VII.*** *All families at the cross-validated optimum, 100 resampled splits, k=6. Differences and bootstrap intervals are quantum minus the row family; a positive difference means the row family is better. Reference baselines are marked* •.

| Dataset | Target | Family | Settings | Test MSE | SE | edf | Q − row | 95% percentile-bootstrap interval |
|---|---|---|---|---|---|---|---|---|
| Tecator | Fat | ARD-RBF | 180 | 0.7025 | 0.0274 | 30.1 | +0.5508 | [+0.432, +0.683] |
| | | • Polynomial (PC) | 10 | 0.7190 | 0.0417 | 27.8 | +0.5342 | [+0.400, +0.676] |
| | | RBF (isotropic) | 180 | 0.7469 | 0.0321 | 33.1 | +0.5064 | [+0.387, +0.641] |
| | | Matérn 5/2 | 180 | 0.7562 | 0.0818 | 47.2 | +0.4971 | [+0.268, +0.690] |
| | | **Quantum** | 180 | **1.2533** | 0.0721 | 27.1 | — | — |
| | | **Twin** | 180 | **1.2789** | 0.0676 | 27.6 | −0.0256 | [−0.087, +0.031] |
| | | PQK | 180 | 2.6621 | 1.0197 | 30.9 | −1.4088 | [−3.586, −0.212] |
| | | • Linear (PC) | 5 | 4.7432 | 0.1128 | 5.8 | −3.4899 | [−3.725, −3.257] |
| | | • Linear (full) | 5 | 5.5258 | 0.2383 | 13.6 | −4.2725 | [−4.779, −3.857] |
| | | • PLS (full) | 300 | 5.8395 | 0.3403 | 10.0 | −4.5862 | [−5.285, −4.001] |
| Tecator | Water | Matérn 5/2 | 180 | 1.8430 | 0.0692 | 58.6 | +1.2288 | [+0.874, +1.627] |
| | | PQK | 180 | 2.7919 | 0.1411 | 31.1 | +0.2799 | [−0.047, +0.633] |
| | | **Twin** | 180 | **2.9514** | 0.1413 | 34.4 | +0.1203 | [−0.131, +0.434] |
| | | **Quantum** | 180 | **3.0718** | 0.1871 | 31.8 | — | — |
| | | RBF (isotropic) | 180 | 3.1144 | 0.5054 | 38.9 | −0.0426 | [−1.185, +0.778] |
| | | ARD-RBF | 180 | 3.2990 | 0.5071 | 45.4 | −0.2273 | [−1.349, +0.624] |
| | | • Polynomial (PC) | 10 | 3.6491 | 0.9033 | 24.3 | −0.5773 | [−2.604, +0.650] |
| | | • Linear (PC) | 5 | 3.8558 | 0.0728 | 5.6 | −0.7840 | [−1.142, −0.393] |
| | | • Linear (full) | 5 | 4.4181 | 0.1772 | 12.3 | −1.3463 | [−1.825, −0.860] |
| | | • PLS (full) | 300 | 4.8895 | 0.3326 | 10.0 | −1.8177 | [−2.535, −1.150] |

| Tecator | Protein | • Linear (full) | 5 | 0.4799 | 0.0147 | 20.4 | $+0.6627$ | $[+0.602, +0.726]$ |
|---|---|---|---|---|---|---|---|---|
| | | • PLS (full) | 300 | 0.5112 | 0.0171 | 13.0 | $+0.6314$ | $[+0.566, +0.698]$ |
| | | Matérn 5/2 | 180 | 0.8908 | 0.0221 | 48.3 | $+0.2518$ | $[+0.193, +0.313]$ |
| | | RBF (isotropic) | 180 | 0.9717 | 0.0295 | 35.0 | $+0.1709$ | $[+0.107, +0.234]$ |
| | | ARD-RBF | 180 | 1.0037 | 0.0292 | 29.2 | $+0.1389$ | $[+0.074, +0.203]$ |
| | | **Twin** | 180 | **1.1355** | 0.0315 | 25.6 | $+0.0071$ | $[-0.013, +0.032]$ |
| | | **Quantum** | 180 | **1.1426** | 0.0323 | 25.7 | — | — |
| | | • Polynomial (PC) | 10 | 1.1483 | 0.0369 | 24.1 | $-0.0057$ | $[-0.072, +0.058]$ |
| | | PQK | 180 | 1.2490 | 0.0692 | 25.0 | $-0.1064$ | $[-0.229, -0.016]$ |
| | | • Linear (PC) | 5 | 1.2922 | 0.0187 | 6.0 | $-0.1496$ | $[-0.216, -0.081]$ |
| Gasoline | Octane | • Linear (full) | 5 | 0.0532 | 0.0020 | 10.4 | $+0.0701$ | $[+0.051, +0.093]$ |
| | | • Linear (PC) | 5 | 0.0621 | 0.0030 | 5.7 | $+0.0612$ | $[+0.042, +0.083]$ |
| | | • PLS (full) | 300 | 0.0714 | 0.0040 | 5.0 | $+0.0520$ | $[+0.032, +0.074]$ |
| | | ARD-RBF | 180 | 0.0737 | 0.0040 | 7.2 | $+0.0497$ | $[+0.031, +0.072]$ |
| | | Matérn 5/2 | 180 | 0.0747 | 0.0042 | 7.5 | $+0.0487$ | $[+0.030, +0.071]$ |
| | | RBF (isotropic) | 180 | 0.0755 | 0.0044 | 7.0 | $+0.0478$ | $[+0.029, +0.070]$ |
| | | **Quantum** | 180 | **0.1233** | 0.0105 | 9.0 | — | — |
| | | **Twin** | 180 | **0.1235** | 0.0105 | 9.1 | $-0.0002$ | $[-0.006, +0.005]$ |
| | | PQK | 180 | 0.1440 | 0.0111 | 17.4 | $-0.0207$ | $[-0.041, -0.002]$ |
| | | • Polynomial (PC) | 10 | 0.1662 | 0.0165 | 21.3 | $-0.0428$ | $[-0.082, -0.008]$ |

As Table VII shows, the quantum kernel ranks fifth of ten on fat, fourth on water, seventh on protein, and seventh on octane; it is best on none. On fat, octane, and protein, the interval against the leading family excludes zero. On water, the quantum kernel is indistinguishable from the isotropic RBF and ARD-RBF families, but Matérn 5/2 attains 1.8430 against 3.0718, an interval that excludes zero comfortably. On protein and octane, a linear ridge regression on the full spectrum is better than the quantum kernel by a factor of more than two.

The twin sits immediately adjacent to the quantum kernel in all four targets, and the paired interval contains zero in each: $-0.0256$ on fat, $+0.1203$ on water, $+0.0071$ on protein and $-0.0002$ on octane. Two further observations strengthen the correspondence beyond what Section 3.4 establishes. First, the two families make the *same hyperparameter choices*: across the 400 target-by-split combinations, the twin and the quantum kernel select the same preprocessing pipeline in 96% of cases (chance would be 20%) and the same bandwidth in 48% of cases (chance 2.8%). Second, they land on the same side of every classical competitor. The twin does not merely match the quantum kernel's error; it reproduces its cross-validation landscape.

Two entries deserve caution. The PQK family on fat has a standard error of 1.0197 against a mean of 2.6621, indicating a heavy-tailed distribution across splits; the mean is not a reliable summary in that cell. On water, ARD-RBF (3.2990) performs slightly worse than the isotropic RBF it nests (3.1144), with standard errors near 0.5 for both; a more flexible family performing worse than the family it contains is more consistent with selection instability than with inductive bias, and we do not read anything into the ordering of those two rows.

A pattern across the table is worth noting. On fat, the second-best family is a degree-two polynomial kernel on the principal component scores, which uses only ten candidate settings and nevertheless outperforms both isotropic RBF and Matérn. Proposition 1 states that the quantum kernel is itself a quadratic form in those same scores. What distinguishes it from the classical quadratic kernels is not the functional form but the weights: the entanglement graph fixes the metric $I + \pi^2 Q$, whereas ARD-RBF and the polynomial kernel are free to tune theirs. On these data, that freedom is worth more than the structure.

## IV. Discussion

### A. The circuit is removable, and we can say what it computed

The central finding is not that a classical kernel happens to perform as well as a quantum one. It is that we can write down, in closed form and without fitting anything, the classical kernel that the ZZ feature map implements in the regime where cross-validation operates. The metric $I + \pi^2 Q$ is read directly off the entanglement graph; the only free quantity is the bandwidth, and it is selected on the same grid as for the quantum kernel.

The evidence that the substitution is faithful goes beyond matched test error. Across 400 target-by-split combinations, the twin and the quantum kernel select the same preprocessing pipeline in 96% of cases, against a chance rate of 20%, and the same bandwidth in 48% of cases, against a chance rate of 2.8%. In the family comparison of Section 3.6, they sit adjacent in all four targets and fall on the same side of every classical competitor. The twin is therefore not an approximation that happens to score similarly; it reproduces the cross-validation landscape of the circuit it replaces.

This distinction matters for how the result should be read. A demonstration that some classical method matches a quantum one leaves open why, and invites the response that a different encoding or a larger device would separate them. An explicit identification of the kernel closes that gap for the encoding studied: the map is a Mahalanobis kernel whose metric is a graph invariant, and any behaviour attributed to it must be a behaviour of that metric.

We also found the surrogate useful as an instrument rather than only as a control. The experiment of Section 3.5.1 required evaluating the kernel at twenty encoding dimensions, which would demand a $2^{20}$-dimensional state vector and is not simulable at the scale of our resampling design. Because the twin costs $\mathcal{O}(k^2)$ per pair, the experiment was routine, and its conclusions were verified against the circuit wherever the circuit could be run. Explicit surrogates make questions about an encoding tractable that the encoding itself puts out of reach.

### B. The phase convention reconciles rather than contradicts

Two conventions for the pair phase circulate, and Proposition 2 shows that the choice is consequential: with the plain product, $\varphi_{ij}(\mathbf{y}) = y_i y_j$ the entangling term enters only at second order in the bandwidth, the induced metric is the identity regardless of the entanglement graph, and the error term is $\mathcal{O}(s^4)$ rather than $\mathcal{O}(s^3)$. The anisotropy of the Qiskit convention comes entirely from its $\pi$-shift, which promotes an entangling contribution from second to first order.

This clarifies rather than overturns the existing picture. Bandwidth-tuned quantum kernels have been reported to resemble radial basis function kernels closely (Flórez-Ablan et al., 2025), with the analysis supported by a model for a separable encoding circuit whose authors note that it captures entangling circuits only qualitatively. Proposition 2 supplies the missing quantitative statement for the convention used in that work: under it, the small-bandwidth kernel *is* an isotropic Gaussian, exactly and independently of entanglement, so the observed resemblance is not a coincidence of the datasets examined but a property of the encoding. Our contribution is to show that the shifted convention is genuinely different (anisotropic, with a metric determined by graph structure) and that this difference, while real, does not translate into predictive advantage.

### C. Entanglement topology as an adjustable inductive bias

Because $M = I + \pi^2 Q$, the eigenvalues of the induced metric are $1 + \pi^2 \mu_i$ with $\mu_i$ the signless Laplacian spectrum of the entanglement graph. The anisotropy of the encoding is thus a graph invariant, and the topology becomes a design parameter with a legible meaning rather than an opaque architectural choice.

Two consequences run against intuition. First, $\mu_1 = 0$ exactly when the graph has a bipartite component, so the path and even cycles leave a direction in which the kernel reduces to the plain Euclidean metric. The most common entanglement pattern is also the one that preserves an isotropic subspace. Second, dense entanglement produces the *least* structured metric: for the complete graph $Q = (k-2)I + J$, so $M$ has only two distinct eigenvalues and is an almost isotropic metric perturbed in a single rank-one direction. If anisotropy is what the encoding contributes, adding entangling gates beyond a sparse graph reduces rather than increases it. Circuit depth acts differently: the interleaved Hadamard layers spread the metric beyond the edges of $G$, so depth widens the interaction range while topology sets its pattern.

### D. Structure is not worth as much as freedom here

The quantum kernel is never the best family in Section 3.6, ranking fifth, fourth, seventh, and seventh of ten. The pattern of what beats it is informative. On the fat target, the second-best family is a degree-two polynomial kernel on the same principal component scores, using ten candidate settings against the quantum kernel's 180. Proposition 1 says the quantum kernel is itself a quadratic form in those scores. The two therefore differ not in functional form but in whose weights they use: the encoding fixes the metric to $I + \pi^2 Q$, while ARD-RBF and the polynomial kernel are free to tune theirs against the data.

Read this way, the encoding supplies a prior (a specific, graph-determined weighting of the principal components and their interactions) and the question is whether that prior is well matched to the problem. On two spectroscopic datasets and four targets, it is not, and a family permitted to learn its metric does better. That is a narrower and more useful statement than "no quantum advantage": it identifies the fixed prior as the binding constraint and suggests that any advantage from such encodings would have to come from choosing a graph whose induced metric is right for the data, which is a classical design question once the correspondence is known.

### E. Selecting the non-classical regime is a symptom, not a signal

Cross-validation chose the regime where the quadratic reduction fails in 15.7% of splits, and forbidding that regime improved test error by 3.8% overall, an effect that a random restriction of equal size does not reproduce. The one exception, the protein target, turned out to have a common cause with the selection behaviour itself. With six principal components, that target is largely unlearnable: ridge regression alone attains 1.58, against 0.45 with twenty components and 0.47 on the full spectrum. The cross-validation objective is correspondingly flat; its two arms differ by 1.3%, and selection is settled by noise. As components are added, the objective sharpens by a factor of 5.6, the arms separate to a ratio of 2.36, and selection of the upper regime falls to zero.

The weaker version of the same effect appears in the other targets at low dimension: fat and water select the upper regime in 37.0% and 46.2% of splits at four components and in 0.0% and 1.6% at twenty. Frequent selection of the saturating regime is thus a symptom of a representation that cannot express the target, not evidence that the regime carries information. This is a usable diagnostic beyond our setting: a quantum kernel whose cross-validated bandwidth drifts toward saturation is a signal to examine the representation before interpreting the kernel.

Raising the dimension is not a general remedy. Fat degrades monotonically beyond six components, and water is flat, then worsening, consistent with the concentration expected at higher encoding dimension (Thanasilp et al., 2024). No single dimension serves all three targets, which is why we treat $k = 6$ as the configuration under study rather than as a recommendation.

### F. Implications for hardware studies and for comparison protocols

Two protocol observations follow from our own experience. Fixing classical length-scale grids independently of the data placed every classical baseline at its grid boundary in every selection, and widening the grids to track the median pairwise distance changed which kernel family appeared to perform best. Similarly, the optimal preprocessing pipeline differs across families within a single target (for protein, the Matérn, RBF and ARD-RBF families select second-derivative preprocessing in 80–94% of splits while the quantum kernel and its twin select raw spectra in 46–47%), so fixing one pipeline handicaps some families and not others. Neither choice is exotic; both are the natural default. Comparisons between quantum and classical kernels are sensitive to them in a way that can change the apparent ordering, and we would encourage reporting the selected values so that boundary effects are visible. A recent hardware diagnostic study of a fixed four-qubit Qiskit ZZ feature map found substantial but incomplete preservation of the statevector kernel geometry, while also emphasizing that implementation fidelity and task relevance are distinct axes (Sipakov, 2026). The configuration studied there fixes the data scaling rather than tuning it, placing the encoding well outside the small-bandwidth regime in which our reduction holds; consistently, the intended statevector kernel is reported to show no label alignment beyond a permutation reference.

## V. Limitations

Several boundaries of the present analysis should be stated plainly.

**The closed form is confined to a single layer.** Proposition 1 gives $M = I + \pi^2 Q$ for $L = 1$. Proposition 3 establishes that the quadratic structure survives at any depth, and Section 3.2 confirms it numerically through $L = 3$ ($R^2 \geq 0.99997$), but for $L \geq 2$ the interleaved Hadamard layers propagate correlations beyond the edges of the entanglement graph and the metric is no longer a function of $Q$ alone. We determine it numerically and leave its closed form open. Practitioners commonly use two layers, so this gap is not incidental.

**The expansion is asymptotic in the bandwidth.** All statements hold to leading order in $s$, with an error term that is $O(s^3)$ under the ZZ convention and tight. The threshold $s^* = 0.072$ at which the quadratic fit falls below $R^2 = 0.99$ is an empirical value for six qubits and our data scaling, not a universal constant. We did not characterise how it moves with the number of qubits, although the derivation suggests it should tighten as the encoding dimension grows.

**Positive definiteness is established only for one layer.** Corollary 1 gives $M \succcurlyeq I \succ 0$ for $L = 1$. For deeper circuits, $M_L$ is a Gram matrix of projected state derivatives and is positive semidefinite but may be singular, in which case the induced form is a pseudometric.

**Two datasets, one domain, one task.** Tecator and Gasoline are standard near-infrared regression benchmarks, but they are two datasets from one measurement modality, and the task throughout is kernel ridge regression. We make no claim about classification, about other spectroscopic modalities, or about data whose structure differs materially from these.

**Six encoding dimensions.** Because our encoding uses one qubit per feature, the number of retained principal components equals the number of qubits, and the main results use six. Section 3.5.1 shows that this choice is near-optimal for two of the three Tecator targets and inadequate for the third, and that no single dimension serves all of them. Results at $k = 6$ should therefore be read as characterising a configuration, not as a recommendation.

**Simulation, not hardware.** All kernels are computed from exact state vectors. Sampling noise, gate error and readout error are absent. Hardware execution introduces an additional discrepancy from the ideal kernel; whether that discrepancy degrades downstream performance or incidentally regularises it is an empirical question we do not address. Hardware effects can also interact with kernel spectra in ways that simulation does not capture.

**The high-dimensional comparison uses the surrogate.** The dimension sweep of Section 3.5.1 evaluates the twin rather than the circuit for $k > 8$, because a twenty-qubit state vector is not simulable at the scale of our resampling design. We verified the substitution where the circuit is computable (the two families agree on the regime choice in 90.6% of splits and on the upper/lower ratio to within 0.4% in median, with the twin systematically about 8% flatter) but the trends reported at $k \geq 10$ are properties of the surrogate.

**Two cells are unstable.** The projected quantum kernel on the fat target has a standard error of 1.0197 against a mean of 2.6621, and on the water target ARD-RBF performs slightly worse than the isotropic RBF it nests, with standard errors near 0.5 for both. We report these as observed and draw no conclusions from their ordering.

## VI. Conclusions

We have shown that the ZZ feature map, in the bandwidth regime that cross-validation selects, induces, to leading order, an anisotropic Mahalanobis geometry with metric $I + \pi^2 Q$ with $Q$ the signless Laplacian of the entanglement graph. The derivation follows from the orthonormality of parity functions and requires no approximation beyond the small-bandwidth expansion; it is verified for path, cycle, and complete graphs to a relative error below $10^{-3}$, and the quadratic structure persists at every circuit depth as a pullback of the Fubini–Study metric. The anisotropy is a consequence of the $\pi$-shifted phase convention: under the unshifted convention, the metric is exactly the identity, which explains why bandwidth-tuned quantum kernels have previously been observed to resemble isotropic radial basis function kernels.

The correspondence is not merely formal. The explicit classical kernel built from this metric requires no fitted parameters and no quantum simulation, yet it matches the quantum kernel's test error in 18 of 20 dataset-by-target-by-preprocessing cells, reproduces its choice of preprocessing pipeline in 96% of splits, and occupies the same position relative to every classical competitor. The one regime in which the reduction fails begins at an identical bandwidth on two dissimilar datasets (consistent with the boundary belonging to the circuit rather than to the data) and forbidding it improves rather than degrades predictive accuracy, by an amount that a random restriction of equal size does not reproduce.

Taken together, these results mean that the quantum circuit can be removed from this pipeline without detectable predictive loss in our experiments, and, more usefully, that we can say what it was computing when it was there. Knowing that a feature map implements a specific graph-determined metric turns questions about the encoding into questions about that metric: which topologies induce which anisotropies, whether a fixed graph-determined prior can be well matched to a problem, and how the induced metric compares with one a classical family is free to learn. On the spectroscopic regression benchmarks studied here, freedom is worth more than the structure the encoding supplies. The quantum kernel is best on none of the four targets, and on two of them, an ordinary ridge regression on the full spectrum is better by more than a factor of two.

We would offer two suggestions for work of this kind. First, classical baselines are sensitive to protocol choices that are easy to make by default: fixed length-scale grids placed every classical baseline at its grid boundary in our setting, and the preferred preprocessing pipeline differs across kernel families within a single target. Reporting selected hyperparameter values makes such boundary effects visible. Second, explicit surrogates are useful beyond validation. Because the twin costs $O(k^2)$ per pair, it allowed us to study the encoding at twenty dimensions, where the circuit itself would require a $2^{20}$-dimensional state vector, and it was there that the one anomaly in our results, a target for which cross-validation persistently selected the non-classical regime, resolved into a property of the representation rather than of the kernel.

**Data Availability**

All code and analysis outputs required to reproduce the results are openly available. The software is archived at Zenodo under DOI 10.5281/zenodo.21978922 and developed at https://github.com/etekeli-cpu/zz-kernel-classical-surrogate. Both datasets are public: Tecator is distributed with the R package **fda.usc** and Gasoline with the R package **pls**; the analysis scripts download them automatically on first run, so no manual retrieval is required. Each table and figure in the paper is produced by a named script listed in the repository README, and all runs are seeded so that results are reproducible to floating point on a given platform.

**Acknowledgements**

The derivations in Section 2.1 and Appendix A were developed and independently verified by the author, who re-derived each result and identified and corrected errors in intermediate drafts. All numerical results were produced by the author's own execution of the deposited code. In the course of this work, the author used Claude (Opus 5, Anthropic) as an assistant for exploratory data analysis, development of the analysis and figure code, drafting and editing of manuscript text, and to support the literature search. All references were located and verified by the author against the original publications. The author takes full responsibility for the content, accuracy, and integrity of the manuscript. This work received no external funding, and the author declares no conflicts of interest.

**References**

Havlíček, V., Córcoles, A.D., Temme, K., Harrow, A.W., Kandala, A., Chow, J.M. & Gambetta, J.M. (2019). Supervised learning with quantum-enhanced feature spaces. Nature, 567, 209–212. https://doi.org/10.1038/s41586-019-0980-2

Schuld, M. & Killoran, N. (2019). Quantum machine learning in feature Hilbert spaces. Physical Review Letters, 122, 040504. https://doi.org/10.1103/PhysRevLett.122.040504

Huang, H.-Y., Broughton, M., Mohseni, M., Babbush, R., Boixo, S., Neven, H. & McClean, J.R. (2021). Power of data in quantum machine learning. Nature Communications, 12, 2631. https://doi.org/10.1038/s41467-021-22539-9

Kübler, J.M., Buchholz, S. & Schölkopf, B. (2021). The inductive bias of quantum kernels. In Advances in Neural Information Processing Systems 34 (NeurIPS 2021), pp. 12661–12673. https://proceedings.neurips.cc/paper/2021/hash/69adc1e107f7f7d035d7baf04342e1ca-Abstract.html

Thanasilp, S., Wang, S., Cerezo, M. & Holmes, Z. (2024). Exponential concentration in quantum kernel methods. Nature Communications, 15, 5200. https://doi.org/10.1038/s41467-024-49287-w

Slattery, L., Shaydulin, R., Chakrabarti, S., Pistoia, M., Khairy, S. & Wild, S.M. (2023). Numerical evidence against advantage with quantum fidelity kernels on classical data. Physical Review A, 107, 062417. https://doi.org/10.1103/PhysRevA.107.062417

Shaydulin, R. & Wild, S.M. (2022). Importance of kernel bandwidth in quantum machine learning. Physical Review A, 106, 042407. https://doi.org/10.1103/PhysRevA.106.042407

Flórez-Ablan, R., Roth, M. & Schnabel, J. (2025). On the similarity of bandwidth-tuned quantum kernels and classical kernels. Quantum Science and Technology, 10(3), 035051. https://doi.org/10.1088/2058-9565/ade7ad

Borggaard, C. & Thodberg, H.H. (1992). Optimal minimal neural interpretation of spectra. Analytical Chemistry, 64(5), 545–551. https://doi.org/10.1021/ac00029a018

Kalivas, J.H. (1997). Two data sets of near infrared spectra. Chemometrics and Intelligent Laboratory Systems, 37(2), 255–259. https://doi.org/10.1016/S0169-7439(97)00038-5

Mevik, B.-H. & Wehrens, R. (2007). The pls package: principal component and partial least squares regression in R. Journal of Statistical Software, 18(2), 1–24. https://doi.org/10.18637/jss.v018.i02

Febrero-Bande, M. & Oviedo de la Fuente, M. (2012). Statistical computing in functional data analysis: the R package fda.usc. Journal of Statistical Software, 51(4), 1–28. https://doi.org/10.18637/jss.v051.i04

Sipakov, R. (2026). Statevector-to-hardware reconstruction of a four-qubit ZZ quantum kernel: a single-backend case study of three execution jobs. *Quantum Reports*, **8**(3), 93. https://doi.org/10.3390/quantum8030093

Schuld, M. (2021). Supervised quantum machine learning models are kernel methods. arXiv:2101.11020v2

Schuld, M., Sweke, R., Meyer, J.J. (2021). Effect of data encoding on the expressive power of variational quantum-machine-learning models. Physical Review A, 103, 032430. https://doi.org/10.1103/PhysRevA.103.032430

Desai, M., Rao, V. (1994). A characterization of the smallest eigenvalue of a graph. Journal of Graph Theory, 18, 181–194. https://doi.org/10.1002/jgt.3190180210

Blank, C., Park, D.K., Rhee, J.-K.K., Petruccione, F. (2020). Quantum classifier with tailored quantum kernel. npj Quantum Information, 6, 41. https://doi.org/10.1038/s41534-020-0272-6

Park, D.K., Blank, C., Petruccione, F. (2020). The theory of the quantum kernel-based binary classifier. Physics Letters A, 384, 126422. https://doi.org/10.1016/j.physleta.2020.126422

Lloyd, S., Schuld, M., Ijaz, A., Izaac, J., Killoran, N. (2020). Quantum embeddings for machine learning. arXiv:2001.03622v2

Huang, H.-Y., Kueng, R., Preskill, J. (2021). Information-theoretic bounds on quantum advantage in machine learning. Physical Review Letters, 126, 190505. https://doi.org/10.1103/PhysRevLett.126.190505

Liu, Y., Arunachalam, S., & Temme, K. (2021). A rigorous and robust quantum speed-up in supervised machine learning. Nature Physics, 17, 1013–1017. https://doi.org/10.1038/s41567-021-01287-z

Jerbi, S., Fiderer, L.J., Poulsen Nautrup, H., Kübler, J.M., Briegel, H.J., Dunjko, V. (2023). Quantum machine learning beyond kernel methods. Nature Communications, 14, 517. https://doi.org/10.1038/s41467-023-36159-y

Cerezo, M., Verdon, G., Huang, H.-Y., Cincio, L., Coles, P.J. (2022). Challenges and opportunities in quantum machine learning. Nature Computational Science 2, 567–576. https://doi.org/10.1038/s43588-022-00311-3

Gil-Fuster, E., Eisert, J., Dunjko, V. (2024). On the expressivity of embedding quantum kernels. Machine Learning: Sci. Technol. 5, 025003. https://doi.org/10.1088/2632-2153/ad2f51

Suzuki, Y., Kawaguchi, H., & Yamamoto, N. (2024). Quantum Fisher kernel for mitigating the vanishing similarity issue. Quantum Science and Technology, 9(3), 035050. https://doi.org/10.1088/2058-9565/ad4b97

Canatar, A., Peters, E., Pehlevan, C., Wild, S. M., & Shaydulin, R. (2023). Bandwidth enables generalization in quantum kernel models. Transactions on Machine Learning Research. https://openreview.net/forum?id=A1N2qp4yAq

Haug, T., Self, C. N., & Kim, M. S. (2023). Quantum machine learning of large datasets using randomized measurements. Machine Learning: Science and Technology, 4, 015005. https://doi.org/10.1088/2632-2153/acb0b4

Peters, E., Caldeira, J., Ho, A., Leichenauer, S., Mohseni, M., Neven, H., Spentzouris, P., Strain, D., Perdue, G.N. (2021). Machine learning of high dimensional data on a noisy quantum processor. npj Quantum Information 7, 161. https://doi.org/10.1038/s41534-021-00498-9

Bowles, J., Ahmed, S., & Schuld, M. (2024). Better than classical? The subtle art of benchmarking quantum machine learning models. arXiv:2403.07059.

Glick, J. R., Gujarati, T. P., Córcoles, A. D., Kim, Y., Kandala, A., Gambetta, J. M., & Temme, K. (2024). Covariant quantum kernels for data with group structure. Nature Physics, 20, 479–483. https://doi.org/10.1038/s41567-023-02340-9

Henry, L.-P., Thabet, S., Dalyac, C., Henriet, L. (2021). Quantum evolution kernel: machine learning on graphs with programmable arrays of qubits. Physical Review A, 104, 032416. https://doi.org/10.1103/PhysRevA.104.032416

Cvetković, D., Rowlinson, P., Simić, S.K. (2007). Signless Laplacians of finite graphs. Linear Algebra and its Applications 423, 155–171. https://doi.org/10.1016/j.laa.2007.01.009

Cvetković, D., & Simić, S. K. (2009). Towards a spectral theory of graphs based on the signless Laplacian, I. Publications de l'Institut Mathématique (Beograd), 85(99), 19–33. https://doi.org/10.2298/PIM0999019C

Cvetković, D., Rowlinson, P., Simić, S.K. (2010). An Introduction to the Theory of Graph Spectra. Cambridge University Press. ISBN 978-0-521-11839-2

Rasmussen, C.E., Williams, C.K.I. (2006). *Gaussian Processes for Machine Learning*. MIT Press. ISBN 026218253X

## APPENDIX A. PROOF OF PROPOSITION 3

Step 1 (Diagonal gate). The decomposition $f_z(sx) = c_z + s\,\gamma_z(x) + s^2\,\eta_z(x)$ yields, via $e^{-i\theta} = 1 - i\theta + O(\theta^2)$,

$$D(sx) = \mathrm{diag}\big(e^{-if_z(sx)}\big)_z = D_0\big(I - is\,\Gamma(x)\big) + O(s^2),$$

where $D_0 = \mathrm{diag}(e^{-ic_z})_z$ and $\Gamma(x) = \mathrm{diag}\big(\gamma_z(x)\big)_z$ is linear in $x$.

Step 2 (L-layer circuit). Writing $A(sx) = D(sx)H^{\otimes k} = A_0 + s\,B(x) + O(s^2)$, where $A_0 = D_0 H^{\otimes k}$ and $B(x) = -i\,D_0\,\Gamma(x)\,H^{\otimes k}$, the circuit $U(sx) = A(sx)^L$ expands as

$$U(sx) = U_0 + s\,V_L(x) + O(s^2), \qquad U_0 = A_0^L,$$

$$V_L(x) = \sum_{\ell=1}^{L} A_0^{L-\ell}\; B(x)\, A_0^{\ell-1}.$$

Each summand inserts $\Gamma(x)$ at a distinct layer position, surrounded by $x$-independent operators; each is linear in $\Gamma(x)$ and hence in $x$. Therefore $V_L(x)$ is linear in $x$.

Step 3 (State). Applying to $|0^{\otimes k}\rangle$,

$$|\phi(sx)\rangle = |\psi_0\rangle + s\,|w(x)\rangle + s^2\,|r(x)\rangle + O(s^3),$$

where $|\psi_0\rangle = U_0|0^{\otimes k}\rangle$ and $|w(x)\rangle = V_L(x)|0^{\otimes k}\rangle$ is linear in $x$. Analyticity of the gate entries in $s$ guarantees a convergent Taylor expansion through $O(s^3)$.

Step 4 (Normalisation). Since $U(sx)$ is unitary, $\langle\phi(sx)|\phi(sx)\rangle = 1$ for all $s$, which forces $\mathrm{Re}\langle\psi_0|w(x)\rangle = 0$.

Step 5 (Fidelity). Let $P = I - |\psi_0\rangle\langle\psi_0|$. Set $|\phi\rangle = |\psi_0\rangle + s|w\rangle + s^2|r\rangle + O(s^3)$ and $|\chi\rangle = |\psi_0\rangle + s|w'\rangle + s^2|r'\rangle + O(s^3)$, and define $\alpha = \langle\psi_0|w\rangle$, $\alpha' = \langle\psi_0|w'\rangle$ (both purely imaginary by Step 4), and $\mu = \alpha' + \bar{\alpha}$. Since $\bar{\alpha} = -\alpha$, we have $\mu = \langle\psi_0|(w' - w)\rangle$, which is again purely imaginary; hence $\mathrm{Re}\mu = 0$ and the $O(s)$ term of $|f|^2$ vanishes. The overlap then reads $f = \langle\phi|\chi\rangle = 1 + s\,\mu + s^2\,C + O(s^3)$, where $C = \langle w|w'\rangle + \langle\psi_0|r'\rangle + \langle r|\psi_0\rangle$, giving

$$|f|^2 = 1 + s^2(2\,\mathrm{Re}(C) + |\mu|^2) + O(s^3).$$

The normalisation constraints yield $2\,\mathrm{Re}(C) = -\|w - w'\|^2$, while $|\mu|^2 = |\langle\psi_0|(w - w')\rangle|^2$. Applying the Pythagorean identity $\|\delta\|^2 = \|P\delta\|^2 + |\langle\psi_0|\delta\rangle|^2$,

$$1 - K(\mathbf{x}, \mathbf{x}') = s^2\|P(|w\rangle - |w'\rangle)\|^2 + O(s^3).$$

Step 6 (Quadratic form). Since $|w(x)\rangle$ is linear in $x$, $|w(x)\rangle - |w(x')\rangle = |w(u)\rangle$ with $u = x - x'$. Writing $|w(u)\rangle = \sum_{i\in V} u_i\,|w(e_i)\rangle$ and using $P = P^\dagger = P^2$,

$$\|P\,|w(\mathbf{u})\rangle\|^2 = \langle w(\mathbf{u})|P|w(\mathbf{u})\rangle = \mathbf{u}^\top M_L\,\mathbf{u},$$

where $(M_L)_{ij} = \mathrm{Re}\langle w(e_i)|P|w(e_j)\rangle$. The matrix $M_L$ is real, symmetric, and positive semidefinite; it depends on the circuit architecture $(U_0, L, G)$ but not on the data.

Two remarks are in order. First, $M_L$ is the Gram matrix of the vectors $P|w(e_i)\rangle$ and is therefore positive semidefinite but not necessarily positive definite: if these vectors are linearly dependent, $M_L$ is singular. The strict positive definiteness established in Corollary 1 for $L = 1$ does not extend automatically to deeper circuits, and the induced form should then be treated as a pseudometric. Second, for $L = 1$ one has $A_0 = D_0 H^{\otimes k}$ and $V_1 = B$, and a direct evaluation returns $M_1 = I + \pi^2 Q$, so Proposition 3 contains Proposition 1 as a special case.

Step 7 (Fubini–Study). The Fubini–Study metric (convention $ds^2_{FS} = \langle d\psi|d\psi\rangle - |\langle\psi|d\psi\rangle|^2$), pulled back to the circuit coordinate $\mathbf{y} = s\mathbf{x}$ via $\mathbf{y} \mapsto \phi(\mathbf{y})$, satisfies

$$g^{(y)}_{ij}(0) = \mathrm{Re}\left[\langle\partial_{y_i}\phi\,\middle|\,\partial_{y_j}\phi\rangle - \langle\partial_{y_i}\phi\,\middle|\,\phi\rangle\langle\phi\,|\,\partial_{y_j}\phi\rangle\right]_{y=0} = (M_L)_{ij}.$$

In the data coordinate $\mathbf{x} = \mathbf{y}/s$, the metric acquires a factor $s^2$: $g^{(x)}_{ij} = s^2\,(M_L)_{ij} + O(s^3)$. Higher-order terms may introduce data dependence; the constant-metric property holds only to leading order.